\documentclass[a4paper,12pt]{article}
\usepackage[round]{natbib}
\usepackage[english]{babel}
\usepackage{setspace}
\usepackage[utf8x]{inputenc}
\usepackage{graphicx} 
\usepackage[left=2cm,right=2cm,top=2cm,bottom=2cm]{geometry}
\usepackage[dvipsnames]{xcolor}
\usepackage{textcomp}
\usepackage[T1]{fontenc}
\usepackage{amssymb}
\usepackage{xcolor}
\usepackage{authblk}
\usepackage{ulem}
\usepackage{hyperref}
\usepackage{lineno}
\usepackage{appendix}
\usepackage{float}
\usepackage{upgreek}
\usepackage{multirow}
\usepackage{booktabs}
\usepackage{lmodern}
\usepackage{url}
\usepackage{soul}
\usepackage{caption,booktabs,array,multirow,makecell,tabularx} 

\title{The impact of the sulfur allotropes and sulfur hydrides on the Venus cloud chemistry}

\author[1]{Maxence Lef{\`e}vre}
\author[1]{Franck Lef\`evre}
\author[1]{Anni M\"a\"att\"anen}
\author[2]{Robert Skog}
\author[2,3]{Benjamin Frandsen}
\author[4]{Aur\'elien Stolzenbach}
\author[5]{Ashwin Braude}

\affil[1]{LATMOS/IPSL, Sorbonne Universit\'e, UVSQ, Universit\'e Paris-Saclay, CNRS, Paris, France}
\affil[2]{University of Helsinki, Faculty of Science, Department of Chemistry, Molecular Science Unit, Finland}
\affil[3]{Tampere University, Faculty of Engineering and Natural Sciences, Aerosol Physics, Finland}
\affil[4]{Instituto de Astrof\'isica de Andaluc\'ia (IAA), Granada, Spain}
\affil[5]{Astera Institute, Emeryville, CA, USA}
\date{Published in JGR: Planets, 131, e2025JE009523. \url{ https://doi.org/10.1029/2025JE009523}}

\begin{document}

\maketitle

\begin{abstract}
Venus is home to vivid sulfur chemistry, with SO$_2$ as the major sulfur gas species and a global cloud layer between 47 and~70 km composed of H$_2$SO$_4$ and H$_2$O. The chemistry in the clouds has been extensively studied with 1D models, but none is able to reproduce the three orders of magnitude decrease of SO$_2$ inside the clouds. Sulfur allotrope chemistry could be a candidate for buffering sulfur atoms, as it can grow into long chains. Sulfur hydrides such as H$_2$S is considered as one of the main sulfur-bearing compounds, possibly linked to surface reactions or have a volcanic origin like on Earth. In this study, we use for the first time a 3D model to study the climatology of the sulfur allotrope in gas and condensed phase and hydride chemistry in the clouds with an explicit photolysis calculation and updated UV cross-section. H$_2$S is close to ppmv levels below the cloud base, the third most abundant sulfur-bearing species in this region after SO$_2$ and OCS. Below the clouds, S$_8$ gas phase is the most abundant sulfur allotrope. Above, the photolysis and condensation are preponderant, and the gas phase decreases to extremely low values. Polysulfur can therefore not be the unknown absorber at cloud-top altitudes. The condensed phase polysulfurs are present from 30 to above 100~km. S$_2$ condensed phase is the most abundant due to its low saturation mixing ratios. Polysulfur species represent a substantial sulfur reservoir in the upper cloud, equivalent to a few ppmv.

\end{abstract}

\section*{Plain Language Summary}
Venus is home to vivid sulfur chemistry, with SO$_2$ as the major sulfur gas species and a 20~km thick global cloud layer composed of H$_2$SO$_4$ and H$_2$O. The cloud chemistry has been extensively studied with 1D models, but none is able to reproduce the SO$_2$ vertical gradient. Sulfur allotropes could be a candidate for buffering sulfur, growing into long chains. Sulfur hydrides such as H$_2$S is considered as one of the main sulfur-bearing compounds, possibly linked to surface reactions or volcanic activity like on Earth. In this study, we use for the first time a 3D model to study the climatology of gas and condensed sulfur allotropes and hydrides in the clouds. H$_2$S is the third most abundant sulfur-bearing species below the cloud base, after SO$_2$ and OCS. Below the clouds, S$_8$ gas phase is the most abundant sulfur allotrope. Above, the gas phase decreases to low values due to photolysis and condensation, and cannot be the unknown absorber at cloud-top altitudes. The condensed polysulfurs are present from 30 km to above 100~km, with S$_2$ as the most abundant due to its low saturation mixing ratio, representing a substantial sulfur reservoir in the upper cloud.

\section{Introduction}

The cloud layer of Venus between 47 to 70~km is home to vivid sulfur chemistry and microphysics, with SO$_2$ as the major sulfur gas species and the condensed phase composed of H$_2$SO$_4$ and H$_2$O. The cloud layer has been extensively observed and modelled. The main discrepancy between the observation and models is the SO$_2$ vertical gradient throughout the clouds \citep{vandaeleSulfurDioxideVenus2017} with three orders of magnitude decrease between the base and top of the cloud layer that no chemical model can reproduce \citep{krasnopolskyPhotochemicalModelVenus2012,biersonChemicalCyclingVenusian2020,rimmerHydroxideSaltsClouds2021,daiInvestigationVenusianAtmospheric2024}.

Spectrophotometer data from VENERA 11, 12, 13 and 14 at 450–600~nm between 10 and 30~km altitude recorded strong absorptions attributed most commonly to gaseous elemental sulfur S$_x$ \citep{maiorovNewAnalysisSpectra2005}, and correlating S$_3$ spectral features from VENERA-11 with an increase with altitude from 0.03~ppbv at 3~km to 0.1~ppbv at 19~km \citep{maiorovNewAnalysisSpectra2005}. The presence of S$_3$ and other sulfur allotropes in the clouds has been hypothesised from Vega probes \citep{porshnevGaschromatographicAnalysisProducts1987}, but no definitive detection of S$_x$ in this region has been made. Polysulfur chemistry could be a candidate for buffering significant sulfur atoms, as it can grow into long chains and condense. Sulfur allotropes have been considered as a candidate for the unknown UV absorber \citep{toonUltravioletAbsorberVenus1982,perez-hoyosVenusUpperClouds2018}.

Sulfur hydrides such as H$_2$S were considered as one of the three main sulfur-bearing compounds with SO$_2$ and OCS \citep{vonzahnCompositionVenusAtmosphere1983}, possibly linked to surface reactions or have a volcanic origin like on Earth. H$_2$S were measured at ppmv levels below the clouds by the Pioneer Venus mission \citep{hoffmanCompositionVenusLower1980,oyamaPioneerVenusGas1980}, and upper limits were determined by SOIR on board Venus Express \citep{mahieuxUpdateSO2Detection2023}. The upcoming DAVINCI mission will measure sulfur allotropes and H$_2$S from the lower cloud to the surface \citep{garvinRevealingMysteriesVenus2022}. \cite{krasnopolskyHighresolutionSpectroscopyVenus2008} estimated an upper limit to the H$_2$S mixing ratio at the cloud tops of 23~ppb with NASA Infrared Telescope Facility ground measurements. 

The chemistry in the Venusian clouds has been extensively studied with 1D models, with some of them featuring polysulfurs and sulfur hydrides. 
The \cite{yungEvidenceCarbonylSulfide2009} 2D model found S$_2$ was the most abundant sulfur allotrope species in the clouds with 0.01~ppmv at 70~km. \cite{krasnopolskyAbundancesImprovedChemical2013} focused on the role of the sulfur allotropes below the clouds using a 1D model, with the sum of all the species reaching 10~ppmv at 45~km. \cite{biersonChemicalCyclingVenusian2020} modelled in 1D the polysulfur chemistry from the surface to 110~km, and found S and S$_2$ as the most abundant species in the clouds with a volume mixing ratio of 10$^{-16}$. None of these models let the allotrope condense. With a 1D model, \cite{zhangSulfurChemistryMiddle2012} quantified the condensation of polysulfurs, S$_8$ solid phase being the most abundant in the clouds with a mixing ratio of 10$^{-12}$.
None of those studies considered the photolysis of S$_8$. \cite{pintoSulfurMonoxideDimer2021} and \cite{frances-monerrisPhotochemicalThermochemicalPathways2022} used 1D models to study the impact of the polysulfur species on the S$_2$O$_2$ chemistry from 58 to 110~km. None of these studies considered the condensation of the sulfur allotropes and photolysis for species with a higher polymerisation degree above 2.

In this study, we use for the first time a 3D model to study the climatology of the sulfur allotrope and hydrides chemistry in the venusian clouds, with an explicit photolysis calculation and updated UV cross-section, to quantify its impact on the cloud sulfur cycle. With many unknowns about the sulfur allotrope chemistry and microphysics, sensitivity tests were conducted on various model parameters.

In Section~\ref{Sec:Model1}, the model is described. In Section~\ref{Sec:H2S}, the sulfur hydrides chemistry is discussed. The impact of polysulfur chemistry is assessed in Section~\ref{Sec:Sx}, and its sensitivity is quantified in Section~\ref{Sec:Sens}. 

\section{Description of the Model Setup}
\label{Sec:Model1}

\subsection{Venus LMDz Planetary Climate Model}

The Venus LMDz Planetary Climate Model (PCM) includes a 3D latitude-longitude grid and physics packages composed of a radiative transfer scheme that accounts for solar heating rates \citep{hausRadiativeHeatingCooling2015}, infrared net exchange rates \citep{eymetNetExchangeParameterization2009} with a cloud model from \cite{hausAtmosphericThermalStructure2014,hausRadiativeHeatingCooling2015}, specific heat changes with temperature \citep{lebonnoisSuperrotationVenusAtmosphere2010}, and coupled photochemistry \citep{stolzenbachThreedimensionalModelingVenus2023,streelStrongVariabilityModeled2026}. The chemical species and cloud particles are not radiatively active. A formalism of level 2.5 of \cite{mellorDevelopmentTurbulenceClosure1982} is adopted as subgrid-scale parameterisation for turbulent mixing, which calculates, with a prognostic equation, the turbulent kinetic energy and mixing length. The Venus LMDz PCM has demonstrated the ability to satisfactorily reproduce zonal super-rotation \citep{garate-lopezLatitudinalVariationClouds2018}. 

\subsection{Chemical Network}

The photochemical model describes the comprehensive chemistries of CO$_2$, CO, hydrogen, oxygen, chlorine, sulfur and nitrogen modeled from the surface to 150~km. The network is built to study chemistry  in the middle atmosphere, down to roughly 35~km, missing reactions favored at high temperature. For solving the ordinary differential equations system of the photochemistry, the Venus LMDz PCM uses the ASIS (Adaptative Semi Implicit Scheme) solver described in detail by \cite{cariolleASISV10Adaptive2017}. The complete list of reactions is available in \citep{stolzenbachThreedimensionalModelingVenus2023,streelStrongVariabilityModeled2026}. 

In this study, we add five species: H$_2$S and its photolysis product HS, and the higher sulfur allotrope polymerisation degrees S$_3$, S$_4$ and S$_8$, as S$_2$ is already present in the network. These additions lead to the inclusion of 32 reactions in the gas-phase chemistry. The sulfur allotropes are considered up to the eighth degree due to their stable ring configuration. To increase the calculation efficiency and with the limited knowledge of these reactions, S$_5$, S$_6$ and S$_7$ are not considered in this study. Additionally, S$_8$ is a good representative of these polysulfur homocycles due to similar molecular and spectral properties \citep{steudelOpticalAbsorptionSpectra1988,trofimovVibrationsSSBond2009,skogUVVisSpectralSimulation2025}. The added reactions are displayed in Table~\ref{T1} with reactions including S$_2$, already implemented in the model.

As for every Venus chemistry model, the SO$_2$ cloud vertical gradient is an issue, and the reference simulations are performed with SO$_2$ set to 10~ppmv in the troposphere to obtain realistic values at the cloud top \citep{krasnopolskyPhotochemicalModelVenus2012,rimmerHydroxideSaltsClouds2021,daiInvestigationVenusianAtmospheric2024}, or set to a realistic value below the clouds leading to inconsistent values at cloud-top \citep{biersonChemicalCyclingVenusian2020}. Simulation with a tropospheric SO$_2$ mixing ratio at 150~ppmv is discussed in the manuscript.

\begin{table}[!ht]
\scriptsize
\label{tab1}
\begin{tabular}{llll}
\hline
ID & Reaction & Rate  & Reference \\
 &  & (in cm$^3$~s$^{-1}$ and cm$^6$~s$^{-1}$  & \\
 & & for two- and three-body reactions)& \\
\hline
1 & S + S + M $\rightarrow$ S$_2$ + M & k$_0$ = 1.19 $\cdot$ 10$^{-29}$ ; k$_\infty$ = 1.0 $\cdot$ 10$^{-10}$ & \cite{nicholasKineticsMechanismDecomposition1979} \\
2 & S + OCS $\rightarrow$ S$_2$ + CO & 
6.63 $\cdot$ 10$^{-20}$ $\cdot$ T$^{2.57}$ $\cdot$ e$^{(-1180/T)}$ & \cite{luExperimentalTheoreticalInvestigation2006} \\
3 & S$_2$ + O $\rightarrow$ SO + S & 2.2 $\cdot$ 10$^{-11}$ $\cdot$ e$^{(-84/T)}$ & \cite{cravenTrajectoryStudiesO21987} \\
\hline
\hline
4 & H$_2$S + OH $\rightarrow$ HS + H$_2$O & 4.0 $\cdot$ 10$^{-12}$ $\cdot$ (T/300)$^{(-2\cdot10^{-1})}$ & \cite{vidalReservoirSulphurDark2017} \\
5 & H + H$_2$S $\rightarrow$ H$_2$ + HS & 1.5 $\cdot$ 10$^{-11}$ $\cdot$  e$^{(-860.0/T)}$ & \cite{vidalReservoirSulphurDark2017} \\

6 & Cl + H$_2$S $\rightarrow$ HCl + HS & 3.7 $\cdot$ 10$^{-11}$ $\cdot$  e$^{(208.0/T)}$ & \cite{atkinsonEvaluatedKineticPhotochemical2004} \\

7 & HS + HS $\rightarrow$ H$_2$S + S & 1.5 $\cdot$ 10$^{-11}$ & \cite{schofieldEvaluatedChemicalKinetic1973}\\

8 & S + H$_2$S $\rightarrow$ HS + HS & 1.7 $\cdot$ 10$^{-10}$ $\cdot$ e$^{(-3620.0/T)}$ & \cite{krasnopolskyChemicalKineticModel2007} \\

9 & HCl + HS $\rightarrow$ Cl + H$_2$S & 7.6 $\cdot$ 10$^{-12}$ $\cdot$ e$^{(-5750./T)}$ & \cite{krasnopolskyChemicalKineticModel2007} \\

10 & HS + H$_2$O $\rightarrow$ H$_2$S + OH & 1.0 $\cdot$ 10$^{-11}$ $\cdot$ e$^{(-14160./T)}$ & \cite{krasnopolskyChemicalKineticModel2007} \\

11 & S + S$_2$ $\rightarrow$ S$_3$ & 6.25 $\cdot$ 10$^{-14}$ $\cdot$ e$^{(450.15*(1/T-1/298.15))}$ & \cite{duKineticsStudyS22011} \\

12 & S$_2$ + S$_2$ $\rightarrow$ S$_3$ + S & 2.8 $\cdot$ 10$^{-11}$ $\cdot$ e$^{(-23000/T)}$ & \cite{krasnopolskyChemicalKineticModel2007} \\

13 & S$_2$ + OCS $\rightarrow$ CO + S$_3$ & 2.4 $\cdot$ 10$^{-11}$ $\cdot$ e$^{(-24900/T)}$ & \cite{krasnopolskyAbundancesImprovedChemical2013} \\

14 & S + S$_3$ $\rightarrow$ S$_2$ + S$_2$ & 1.7 $\cdot$ 10$^{-10}$ $\cdot$ e$^{(-2800/T)}$ & \cite{krasnopolskyChemicalKineticModel2007} \\

15 & CO + S$_3$ $\rightarrow$ S$_2$ + OCS & 1.0 $\cdot$ 10$^{-11}$ $\cdot$ e$^{(-20000/T)}$ & \cite{krasnopolskyAbundancesImprovedChemical2013} \\

16 & H + S$_3$ $\rightarrow$ HS + S$_2$ & 1.2 $\cdot$ 10$^{-10}$ $\cdot$ e$^{(-1950/T)}$ & \cite{krasnopolskyAbundancesImprovedChemical2013} \\

17 & S + S$_3$ + M $\rightarrow$ S$_4$ + M & k$_0$ = 1.0 $\cdot$ 10$^{-30}$ ; k$_\infty$ = 3.0 $\cdot$ 10$^{-11}$ & \cite{nicholasKineticsMechanismDecomposition1979,millsObservationsPhotochemicalModeling1998} \\

18 & S$_2$ + S$_2$ + M $\rightarrow$ S$_4$ + M & k$_0$ = 2.2 $\cdot$ 10$^{-29}$ ; k$_\infty$ = 1.0 $\cdot$ 10$^{-10}$ & \cite{nicholasKineticsMechanismDecomposition1979,millsObservationsPhotochemicalModeling1998} \\

19 & S$_4$ + S$_4$ + M $\rightarrow$ S$_8$ + M & k$_0$ = 1.0 $\cdot$ 10$^{-30}$ ; k$_\infty$ = 3.0 $\cdot$ 10$^{-11}$ & \cite{nicholasKineticsMechanismDecomposition1979,millsObservationsPhotochemicalModeling1998} \\ 

20 & S$_3$ + O $\rightarrow$ SO + S$_2$ & 2.0 $\cdot$ 10$^{-11}$ $\cdot$ e$^{(-500./T)}$ & \cite{mosesPostSL9SulfurPhotochemistry1995,zahnlePhotolyticHazesAtmosphere2016} \\

21 & S$_4$ + O $\rightarrow$ SO + S$_3$ & 2.0 $\cdot$ 10$^{-11}$ $\cdot$ e$^{(-500./T)}$ & \cite{mosesPostSL9SulfurPhotochemistry1995,zahnlePhotolyticHazesAtmosphere2016} \\
22 & SO$_3$ + CO $\rightarrow$ SO$_2$ + CO$_2$ & 1.0 $\cdot$ 10$^{-11}$ $\cdot$ e$^{(-13000./T)}$ & \cite{krasnopolsky2OH2SOSystem1994} \\
23 & SO$_2$ + CO $\rightarrow$ SO + CO$_2$ & 4.5 $\cdot$ 10$^{-12}$ $\cdot$ e$^{(-24300./T)}$ & \cite{bauerReactionCOSO21971,krasnopolskyChemicalKineticModel2007} \\
24 & S$_2$ + CO $\rightarrow$ OCS + S & 1.0 $\cdot$ 10$^{-12}$ $\cdot$ e$^{(-17460/T)}$ & \cite{krasnopolskyChemicalKineticModel2007} \\
25 & H + OCS $\rightarrow$ CO + HS & 1.2 $\cdot$ 10$^{-11}$ $\cdot$ e$^{(-1950/T)}$ & \cite{leeAbsoluteRateParameters1977} \\
27 & S + SO$_2$ $\rightarrow$ SO + SO & 2.3 $\cdot$ 10$^{-11}$ $\cdot$ e$^{(-5200/T)}$ & \cite{krasnopolskyChemicalKineticModel2007} \\
28 & HS + CO $\rightarrow$ OCS + H & 4.2 $\cdot$ 10$^{-14}$ $\cdot$ e$^{(-7660/T)}$ &  \cite{kurbanovRoleReactionCO1995} \\
29 & SO$_2$ + H $\rightarrow$ SO + OH & 3.7 $\cdot$ 10$^{-9}$ $\cdot$ e$^{(-14350/T)}$ & \cite{krasnopolskyChemicalKineticModel2007} \\
30 & OCS + OH $\rightarrow$ CO$_2$ + HS & 1.1 $\cdot$ 10$^{-13}$ $\cdot$ e$^{(-1200/T)}$ & \cite{atkinsonEvaluatedKineticPhotochemical2004} \\
31 & S + S$_4$ $\rightarrow$ S$_2$ + S$_3$ & 4.0 $\cdot$ 10$^{-11}$ $\cdot$ e$^{(-500/T)}$ & \cite{mosesPostSL9SulfurPhotochemistry1995,zahnlePhotolyticHazesAtmosphere2016} \\
32 & H$_2$ + HS $\rightarrow$ H + H$_2$S & 6.52 $\cdot$ 10$^{-12}$ $\cdot$ (T/300)$^{(9.00\cdot 10^{-2)}}$ $\cdot$ e$^{(-8050/T)}$ & \cite{leenOxygenChemistryShocked1988} \\ 
33 & HS + CO$_2$ $\rightarrow$ OCS + OH & 2.6 $\cdot$ 10$^{-13}$ $\cdot$ e$^{(-19360/T)}$ & \cite{krasnopolskyChemicalKineticModel2007} \\
34 & S$_4$ + H $\rightarrow$ S$_3$ + HS & 5.0 $\cdot$ 10$^{-11}$ $\cdot$ e$^{(-500/T)}$ & \cite{zahnlePhotolyticHazesAtmosphere2016} \\ 
35 & HS + HS $\rightarrow$ H$_2$ + S$_2$ & 3.0 $\cdot$ 10$^{-11}$ $\cdot$ e$^{(-20600/T)}$ & \cite{zahnlePhotolyticHazesAtmosphere2016}\\
\hline
\end{tabular}
\caption{List of the H$_2$S and S$_x$ reactions in the Venus LMDz PCM. The first three reactions were already present in the model.}
\label{T1}
\end{table}

\subsection{Photodissociation Scheme}

In this study, the photolysis rates are calculated online, following the same formalism as in the Mars version of the LMDz PCM \citep{lefevreRelationshipOzoneWater2021}. The possible screening effect of all ultraviolet-absorbing species in the computation of the photolysis rates is taken into account. The radiative transfer computation is performed over the range of 0-815~nm in 194 spectral bins, with a variable resolution adapted to resolve the Lyman-alpha line (0.2~nm) and the fine structures of the CO$_2$ and SO$_2$ absorption cross-sections (0.5-1.0~nm). The model's complete list of photolytic processes is
available in \cite{stolzenbachThreedimensionalModelingVenus2023,streelStrongVariabilityModeled2026}.

With the addition of the five new gas species, four photodissociation reactions were added. These reactions are displayed in Table~\ref{T2}. 

\begin{table}[!ht]
\label{tab2}
\begin{tabular}{lll}
\hline
ID & Photodissociation & cross-section Reference \\
\hline
1 & H$_2$S + h$\upnu$ $\rightarrow$ HS + H & \cite{wuTemperaturedependentPhotoabsorptionCross1998,burkholderChemicalKineticsPhotochemical2020}  \\
2 & S$_2$ + h$\upnu$ $\rightarrow$ S + S & \cite{starkFouriertransformspectroscopicPhotoabsorptionCross2018} + \cite{skogUVVisSpectralSimulation2025} \\
3 & S$_3$ + h$\upnu$ $\rightarrow$ S$_2$ + S & \cite{skogUVVisSpectralSimulation2025} \\
4 & S$_4$ + h$\upnu$ $\rightarrow$ S$_2$ + S$_2$ & \cite{skogUVVisSpectralSimulation2025} \\
5 & S$_8$ + h$\upnu$ $\rightarrow$ S$_4$ + S$_4$ & \cite{skogUVVisSpectralSimulation2025} \\
\hline
\end{tabular}
\caption{List of the added Photodissociation in the Venus PCM}
\label{T2}
\end{table}

The S$_2$ UV cross-section used previously by the Venus PCM was based on \cite{millsObservationsPhotochemicalModeling1998} estimation. Previous 1D model \citep{krasnopolskyPhotochemicalModelVenus2012,zhangSulfurChemistryMiddle2012,biersonChemicalCyclingVenusian2020} used 
\cite{billmersUltravioletvisibleAbsorptionSpectra1991} as UV cross-section for S$_3$ and S$_4$. No Venus studies considered the photolysis of S$_8$. For S$_8$ in an exoplanet context, \cite{zahnlePhotolyticHazesAtmosphere2016} considered the possibility that a UV photon absorbed would break the ring, and then the linear S$_8$ molecule would either be put back into a ring or split into two S$_4$ molecules by a secondary photon in the visible spectral range using cross-section from \citep{meyerElementalSulfur1976}. This mechanism is discussed in Section~\ref{S8}.

\begin{figure}[!ht]
  \centering
  \includegraphics[width=\textwidth]{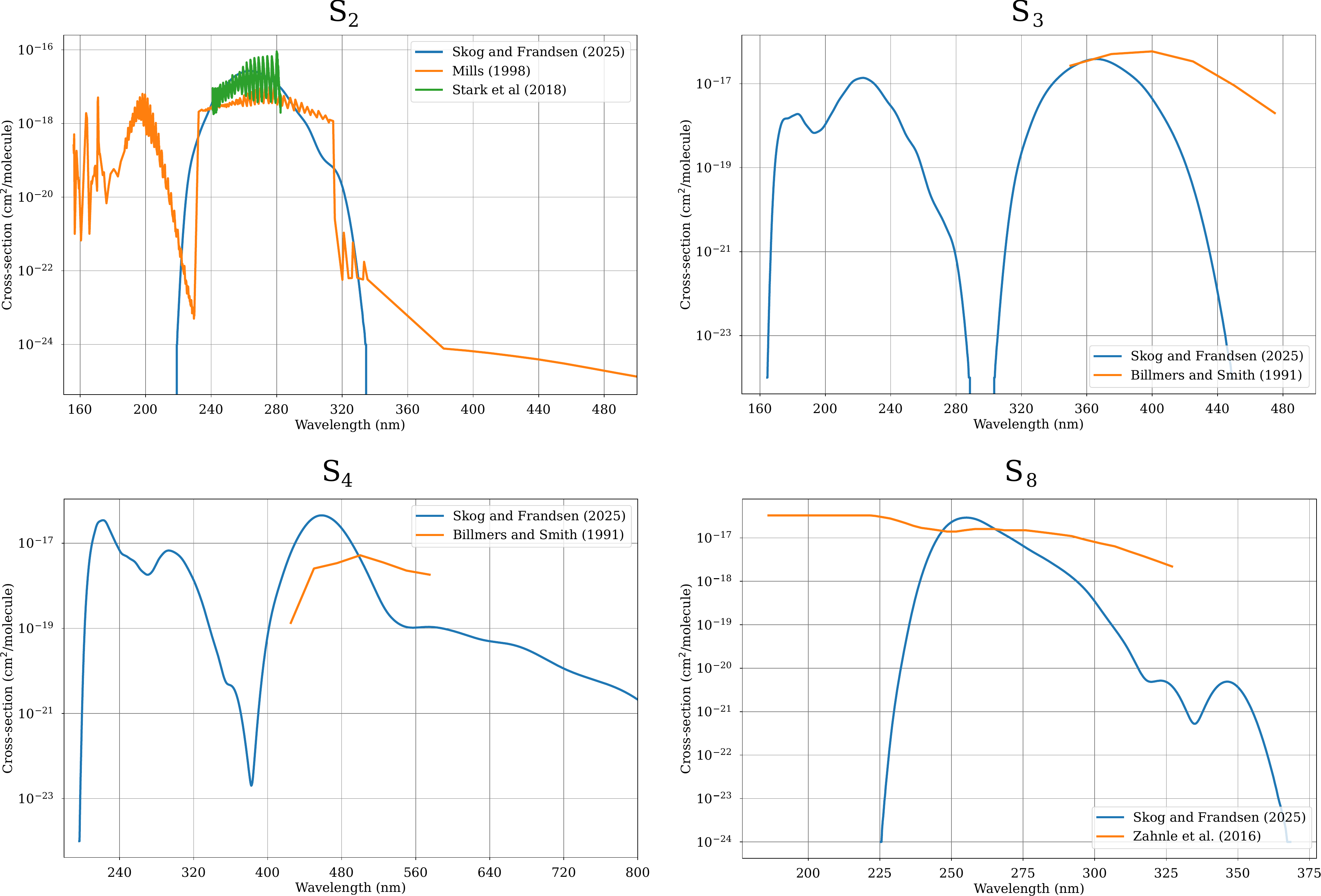}
    \caption{Cross-section of S$_2$, S$_3$, S$_4$ and S$_8$ gas in the UV and visible used so far in different chemical models (orange curves) and calculated for this study (blue curves). The green line corresponds to recent laboratory measurements of S$_2$ cross-section \citep{starkFouriertransformspectroscopicPhotoabsorptionCross2018}.}
  \label{231}
\end{figure}

To accurately model the photodissociation of the different sulfur allotropes, recently published, updated cross-sections were used \citep{skogUVVisSpectralSimulation2025}. These cross-sections were calculated using the nuclear ensemble approximation approach, an approach that has also been utilized to calculate cross-sections for other molecules believed to be important in the Venusian atmosphere \citep{farahaniSimulatedElectronicAbsorption2019,frandsenSpectroscopyOSSOOther2020,trabelsiOCS2IsomersVenusian2024,chaoUVVisSpectrum2025,trabelsiSpectroscopicCharacterizationS3O2025}. A brief description of the method, as it pertains to the cross-section calculations for the sulfur allotropes, is provided in the Supporting Information (SI). Figure~\ref{231} shows different cross-sections for S$_2$, S$_3$, S$_4$ and S$_8$ gas. The orange lines represent the cross-section used in previous studies, and the blue lines represent the new calculations. The green line is the recent experimental measurements for S$_2$ \citep{starkFouriertransformspectroscopicPhotoabsorptionCross2018}, in agreement with the updated cross-section.  

S$_4$ has multiple stable isomers with different UV cross-sections. The linear-cis isomer is the one with the lowest energy \citep{quelchResolutionLongstandingProblem1990}, and for simplicity, only this one is considered in this study. 

\begin{figure}[!ht]
  \centering
  \includegraphics[width=10cm]{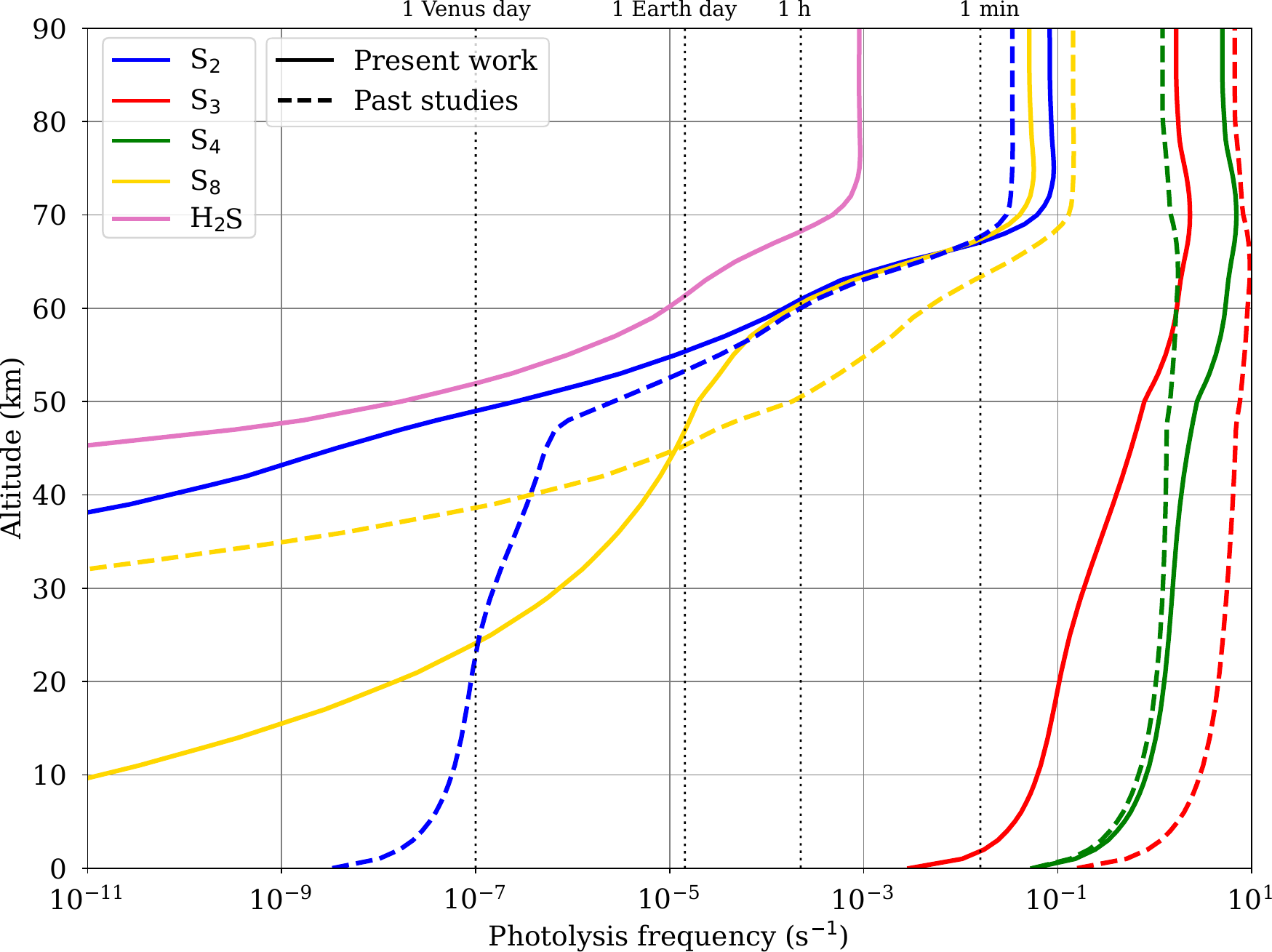}
    \caption{Photolysis rate vertical profile (log$_{10}$ s$^{-1}$) at the Equator at noon for S$_2$, S$_3$, S$_4$, S$_8$ and H$_2$S gas phase with the UV cross-section corresponding to orange curves in Figure~\ref{231} are the dashed curves, and updated UV cross-section corresponding to blue lines in Figure~\ref{231} are the solid curves. The vertical dotted curves correspond to the photolysis timescale.}
  \label{232}
\end{figure}

Major differences can be observed between the different sets of cross-sections for all the species. For S$_2$, there is no more absorption at visible wavelengths. For S$_3$ and S$_4$, a strong UV absorption is now occurring. For S$_8$, the spectral range is shifted towards longer wavelength with a smaller amplitude for the updated cross-sections. Figure~\ref{232} shows the photolysis vertical profiles in log$_{10}$ s$^{-1}$ at the Equator at noon for the different sulfur allotrope species with the UV cross-sections used in previous studies (orange lines in Figure~\ref{231}) and the ones used in this study (blue lines in Figure~\ref{231}). For S$_2$, the photolysis rates are comparable from 90~km to 50~km, but below the updated rate decreases faster due to the absence of absorption in the visible range. The updated S$_3$ photolysis rate has a similar decrease with altitude compared to the rate with the previous cross-section, only divided by a ratio between 10 and 100. For S$_4$, the updated photolysis rate above 50 km is much greater than the rate with the previous cross-sections, while they are not very different below 40 km. This is possibly mainly due to the presence of absorption in the UV range for the new cross-section compared to the previous one (bottom left panel of Fig~\ref{231}). The updated S$_8$ photolysis rate has a similar decrease with altitude compared to the rate with the previous cross-section until 60~km, but decreased by a factor of 10. Below, the previous cross-section decreases significantly faster than the updated one, due to the absorption above 320~nm. Inside the cloud layer, the photolysis timescale for S$_8$ drops from less than one minute to 1 Earth day at cloud bottom. For S$_3$ and S$_4$, and for S$_2$ and S$_8$ above 60~km, photolysis is the primary chemical sink for the sulfur allotropes.

\subsection{Condensation Scheme}

The PCM takes into account the condensation of H$_2$SO$_4$ and H$_2$O based on the hypothesis that clouds are in a state of thermodynamic equilibrium, i.e. following exactly the saturation pressure profile of the calculated equilibrium H$_2$SO$_4$ aqueous solution. The cloud particle size distributions for the three modes implemented are prescribed and are the same as in Table 5 from \cite{stolzenbachThreedimensionalModelingVenus2023}.

\begin{figure}[!ht]
  \centering
  \includegraphics[width=10cm]{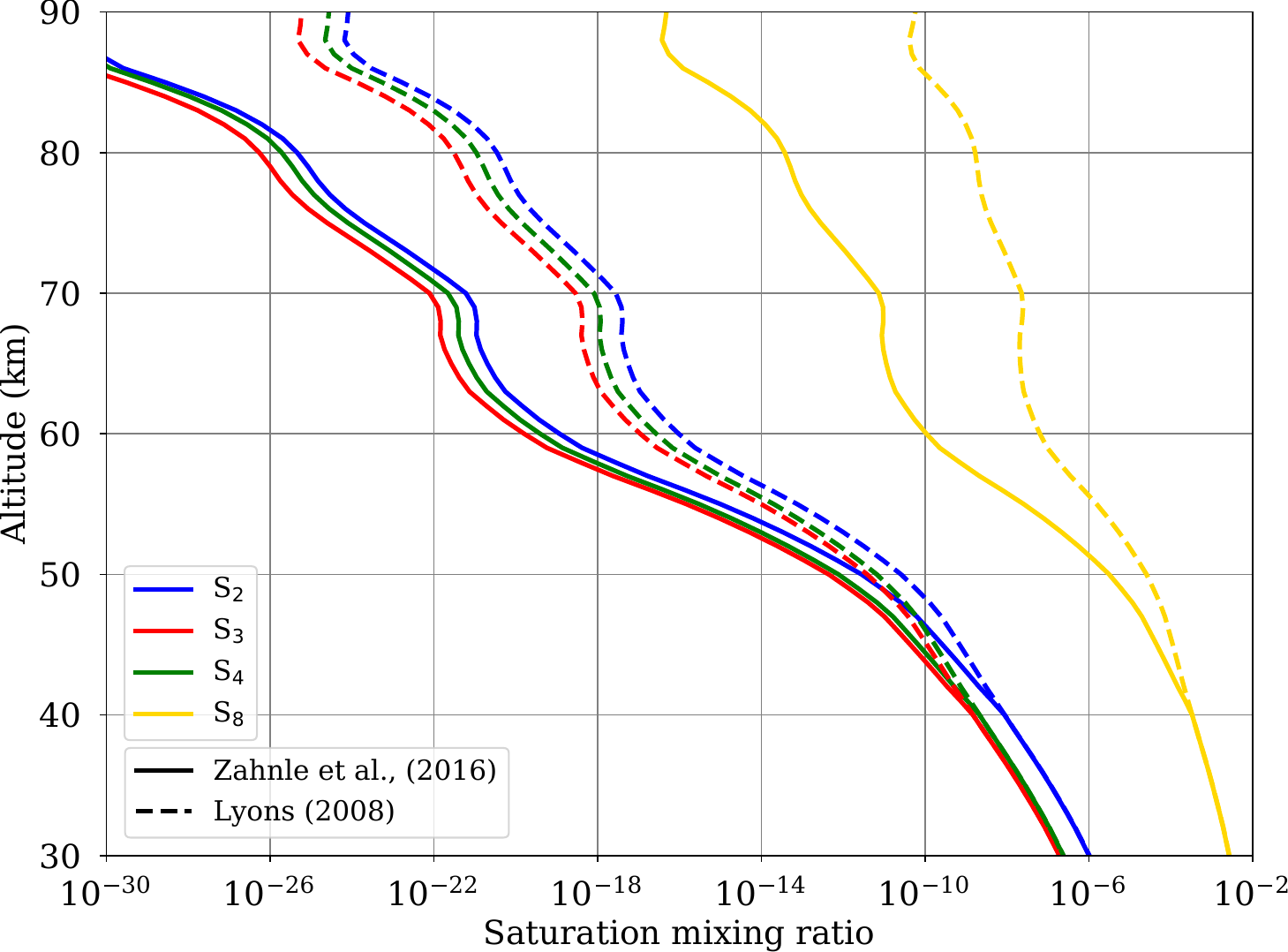}
    \caption{Saturation mixing ratio for S$_2$, S$_3$, S$_4$ and S$_8$ from \cite{zahnlePhotolyticHazesAtmosphere2016} (solid curves) and  \cite{lyonsEstimateEquilibriumSpeciation2008} (dashed curves).}
  \label{241}
\end{figure}

The sulfur allotropes from S$_2$ to S$_8$ are thought to condense in the Venusian conditions \citep{millsAtmosphericCompositionChemistry2007,zhangSulfurChemistryMiddle2012}. The same philosophy as for H$_2$SO$_4$ aqueous solution is used for the polysulfur condensation/evaporation in this study. Two sets of saturation mixing ratios are available in the literature, as shown in Figure~\ref{241}: \cite{lyonsEstimateEquilibriumSpeciation2008} and \cite{zahnlePhotolyticHazesAtmosphere2016}, differentiating in the hypothesis made at low temperature. The saturation curves of \cite{zahnlePhotolyticHazesAtmosphere2016} are used as the reference in this study, and the difference with \citep{lyonsEstimateEquilibriumSpeciation2008} is discussed in Section~\ref{condensation}.

The radius of condensed sulfur allotropes is not known, but would be under 1~$\upmu$m \citep{onischukSurfaceTensionSulfur2016}. In this work, the radius is fixed to 0.1~$\upmu$m for the four condensed species, on the order of the smallest cloud particles \citep{titovCloudsHazesVenus2018}. A sedimentation flux is also taken into account, terminal velocity $V$ based on the Stokes law \citep{fuchsMechanicsAerosols1964,ackermanPrecipitatingCondensationClouds2001}: 

\begin{equation}
    V = \frac{2\upbeta r^2 g(\uprho_p - \uprho)}{9\upeta}
\end{equation}

With $r$ is the particle radius, $g$ the gravitational acceleration, $\uprho_p$ is the particle density, $\uprho$ is the atmosphere density, $\upeta$ is the viscosity of the atmospheric gas calculated following \citep{johnstonViscositiesCarbonMonoxide1942} and the \cite{jonesDeterminationMolecularFieldsI1924} formalism, and $\upbeta$ is the Cunnigham slip factor given by \citep{fuchsMechanicsAerosols1964}:
\begin{equation}
    \upbeta = 1 + K_n(1.256 + 0.4e^{-1.1/Kn})
\end{equation}

With $K_n$ is the Knudsen number, equal to the ratio of the mean free path $\uplambda$ to the size of radius $r$ of the particle: $K_n$ = $\uplambda$/r

The condensed particle mean free path $\uplambda$ is given by
\begin{equation}
    \uplambda = \frac{k_B T}{\sqrt{2}\pi d}\frac{1}{P}
\end{equation}

With $k_B$ the Boltzmann constant, $T$ and $P$ the temperature and pressure of the gas and
$d$ is the molecular diameter.

The value of these different physical parameters for three altitudes in the clouds and particle radius of 0.1 and 1~$\upmu$m is displayed in Table~S2. The sensitivity to the particle radius will be discussed in Section~\ref{radius}. There is no interaction between the different condensed allotropes, nor with sulfuric acid and water droplets. The condensed sulfur allotropes do not take part in any heterogeneous chemical reaction with any gas species. 

\subsection{Simulation Settings}

We use the configuration of the Venus PCM with 78 vertical levels, extending up to approximately 150~km in the thermosphere \citep{gilliThermalStructureUpper2017}. The vertical layers of the model are displayed in Table~\ref{ST2}. This version includes parameterisations of CO$_2$-O cooling, non-LTE CO$_2$ heating, thermal conduction, molecular viscosity, extreme ultraviolet heating, and non-orographic gravity waves \citep{gilliVenusUpperAtmosphere2021}. The horizontal resolution of the simulations presented here is set to 3.75$^{\circ}$$\times$1.875$^{\circ}$ (96 longitudes$\times$96 latitudes). The simulations are run for 9 Venus days with the added species, long enough for SO$_2$ and polysulfur to reach equilibrium in the clouds.

\begin{figure}[!ht]
  \centering
  \includegraphics[width=10cm]{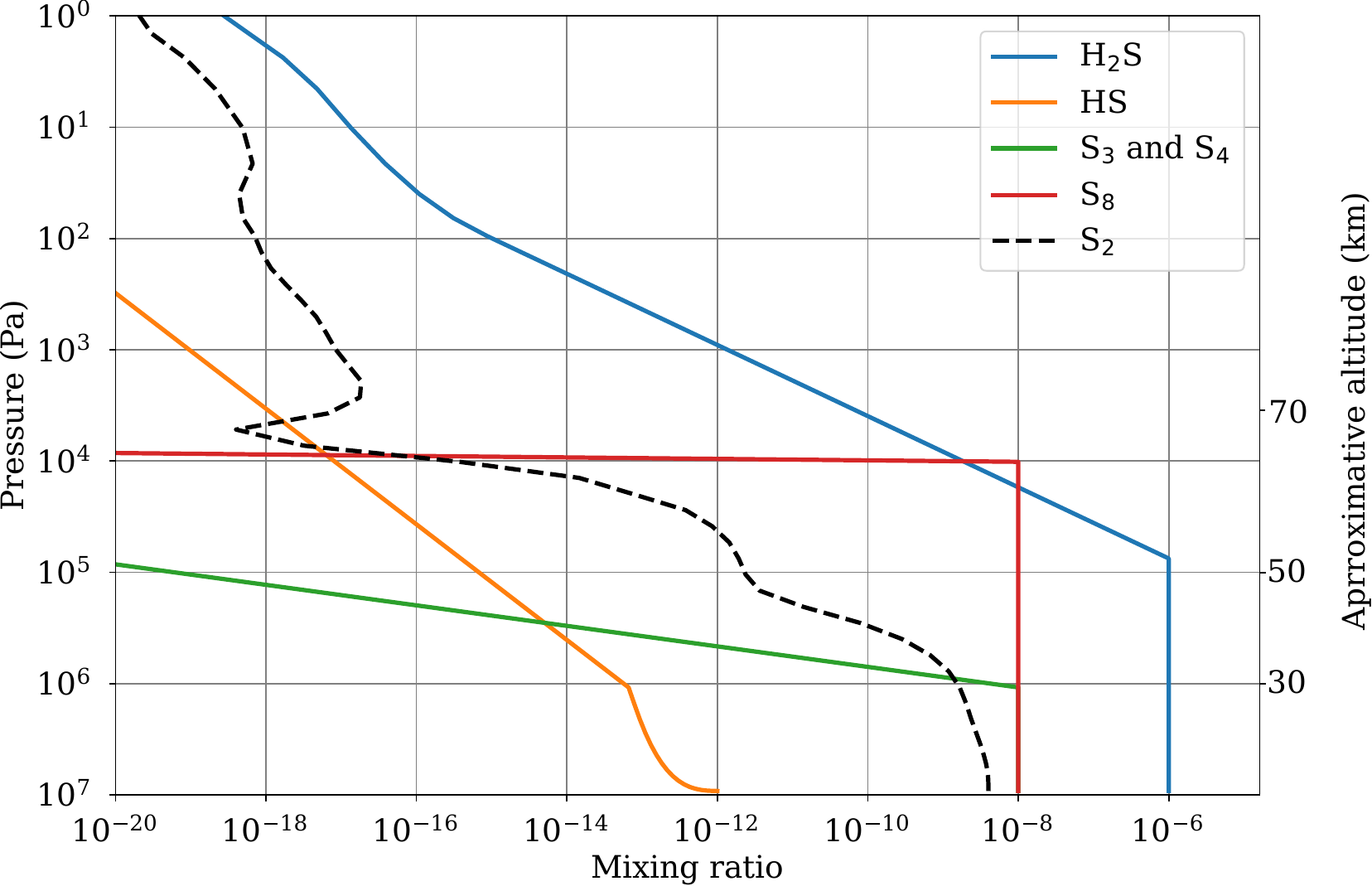}
    \caption{Initialisation vertical profiles of the five new gas species (plain curves) mixing ratio and S$_2$ in gas phase horizontally averaged (dashed curve) already present in the model.}
  \label{251}
\end{figure}

The initial vertical profile uniformly imposed in the horizontal direction of the five added gaseous species is shown in Figure~\ref{251}. These gaseous species are initialised below the cloud to the order of magnitude measured or estimated from past modelling \citep{johnsonVenusAtmosphericComposition2019,maiorovNewAnalysisSpectra2005,krasnopolskyAbundancesImprovedChemical2013} and then decrease to 10$^{-30}$ at the top of the model for the sulfur hydrides and to zero for the sulfur allotropes in the clouds. The condensed phase of sulfur allotropes is initialised at zero everywhere. The other gaseous species are initialized from the reference simulation of \citep{stolzenbachThreedimensionalModelingVenus2023}.

\section{Sulfur Hydrides Chemistry}
\label{Sec:H2S}

\begin{figure}[!ht]
  \centering
  \includegraphics[width=\textwidth]{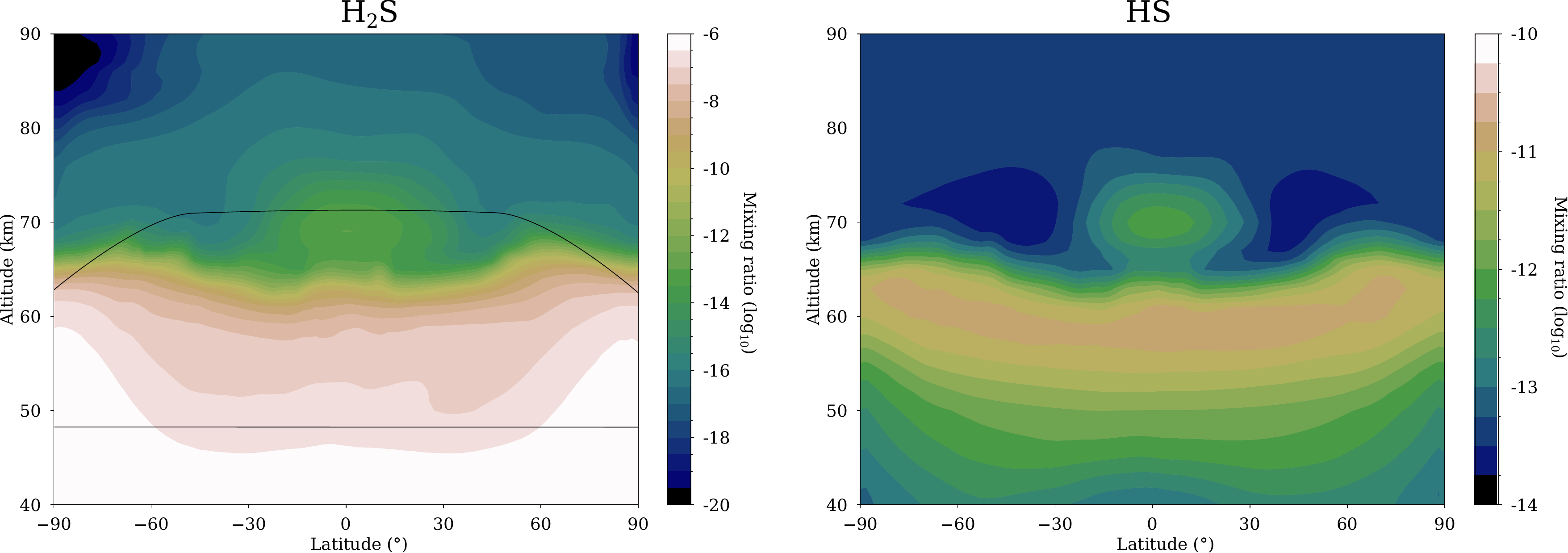}
    \caption{Instantaneous zonally-averaged vertical cross-section of H$_2$S (left) and HS (right). The black contour on the left panel corresponds to the bottom and top altitudes of the clouds \citep{ignatievAltimetryVenusCloud2009, titovCloudsHazesVenus2018}.}
  \label{302}
\end{figure}

Figure~\ref{302} shows the instantaneous zonally-averaged maps of H$_2$S and HS. H$_2$S is close to the ppmv level up to 50~km near the equator and up to 60~km at high latitudes to the pole. It then slightly decreases up to roughly 65~km, and strongly decreases above. In contrast, for HS, a 10~km thick layer with a mixing ratio of the order of 100~pptv is present between 55 and 65~km of altitude. This layer is due to the photolysis of H$_2$S, efficient enough to prevent the vertical mixing of tropospheric H$_2$S-rich parcels to cloud-top altitudes by the general circulation. Sulfur hydrides represent a sulfur sink of approximately 1~ppmv in the lower cloud. The values in the upper cloud are consistent with the upper-limits varying between 1~pptv and 100~ppmv at 60~km determined by SOIR \citep{mahieuxUpdateSO2Detection2023}. At cloud top altitudes, H$_2$S mixing ratio is below the 23~ppbv upper-limit calculated with ground measurements \citep{krasnopolskyHighresolutionSpectroscopyVenus2008}. With a tropospheric SO$_2$ mixing ratio of 150~ppmv, H$_2$S mixing ratio is at the ppmv level up to 60~km for all latitudes. The maximum mixing ratio of HS increased by one order of magnitude. Below the clouds, H$_2$S is stabilising around 1~ppmv, the third most abundant sulfur-bearing species after SO$_2$ and OCS.

\section{Sulfur Allotropes Chemistry}
\label{Sec:Sx}

\begin{figure}
  \centering
  \includegraphics[width=14cm]{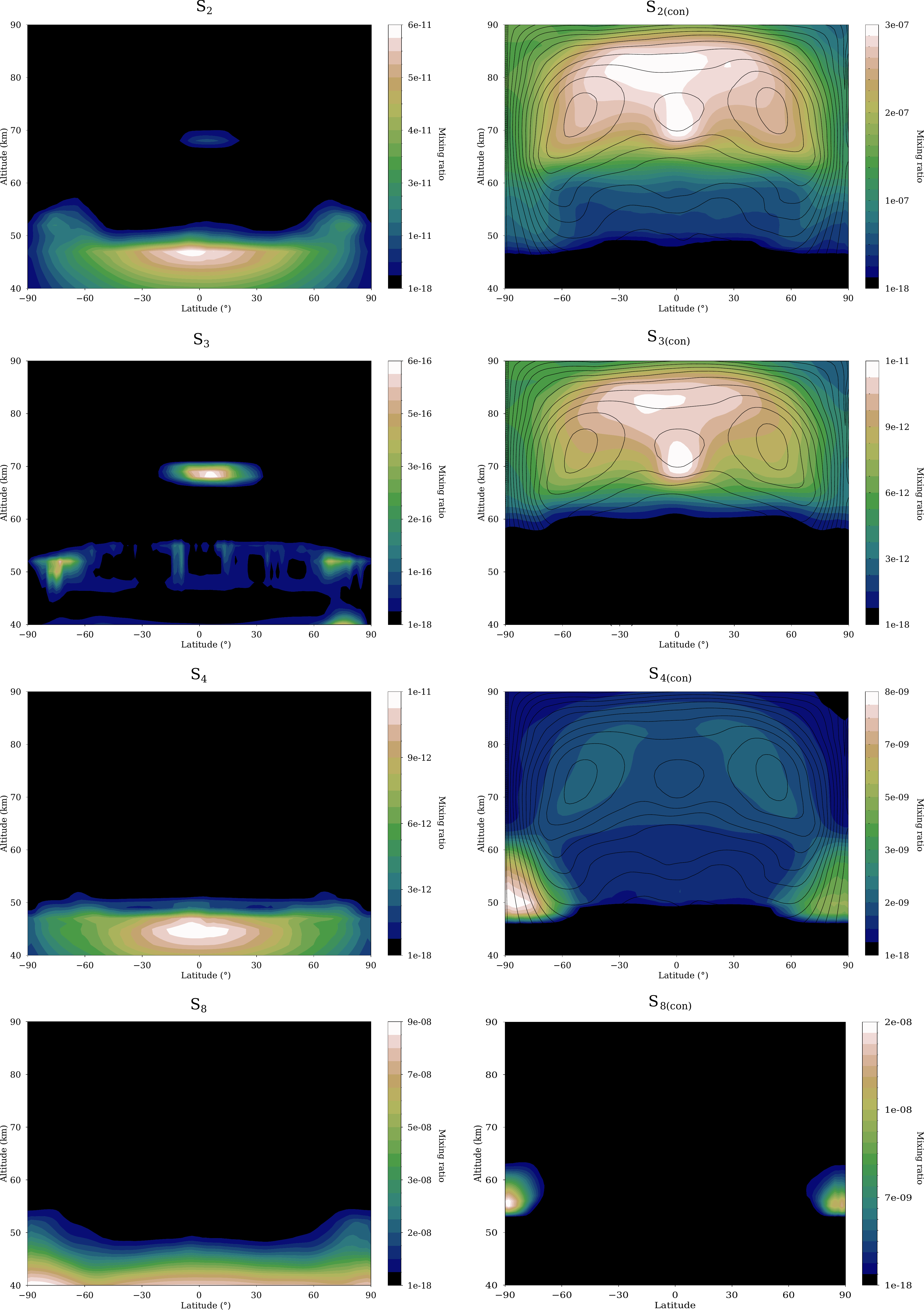}
    \caption{Instantaneous zonal-averaged maps of the sulfur allotropes gas (left) and condensed (right) phase mixing ratio. Mixing ratios can reach values below the 10$^{-18}$. Black contours are the zonal averaged of the zonal wind representing the general circulation.}
  \label{311}
\end{figure}

Figure~\ref{311} shows the instantaneous zonally-averaged maps of S$_2$, S$_3$, S$_4$ and S$_8$ in both the gas (left) and condensed (right) phase. The gas phase of the four considered sulfur allotropes is present mainly below 55~km with S$_8$ as the main species. Above, the photolysis and/or condensation reduce their lifetime, and their volume mixing ratio is typically below 10$^{-16}$. Therefore, sulfur allotropes are not viable candidates for the unknown UV absorber as previously speculated \citep{toonUltravioletAbsorberVenus1982}.

The condensed phase of these species is present from 30 to above 100~km, i.e. with a mixing ratio above 10${-30}$, the altitude range of the lower and upper haze \citep{titovCloudsHazesVenus2018}, and could serve as precursors of aerosols but not as the main species as previously speculated \citep{bertauxActiveSpectrometryUltraviolet1986}. The sulfur allotropes in the condensed phase could also play a role as cloud condensation nuclei for the sulfuric acid and water droplets. Their peak abundance is around 70~km for S$_2$ and S$_3$, at 50~km for S$_4$ and at 55~km for S$_8$. S$_2$, S$_3$ and S$_4$ in condensed phase are present at all latitudes, dominated by the super-rotation wind pattern, the equatorial and mid-latitudes jets in the 60-80~km region. S$_2$ in condensed phase is the dominant species due to its lower saturation mixing ratio and the availability of gas phase sulfur atoms and S$_2$ as a photolysis product (Table~\ref{T2}). S$_4$ is maximum above 60$^{\circ}$ of latitude, and S$_8$ is only present above 70$^{\circ}$ of latitude. This high-latitude bias for higher polymerisation degrees is due to the lower photolysis rate, allowing allotropes to grow to the 8th indice and condense, and the Hadley cell brings S$_2$ condensed-rich parcels to higher saturation mixing ratios, and thus contributes to building larger degrees of allotropes.

\begin{figure}[!ht]
  \centering
  \includegraphics[width=\textwidth]{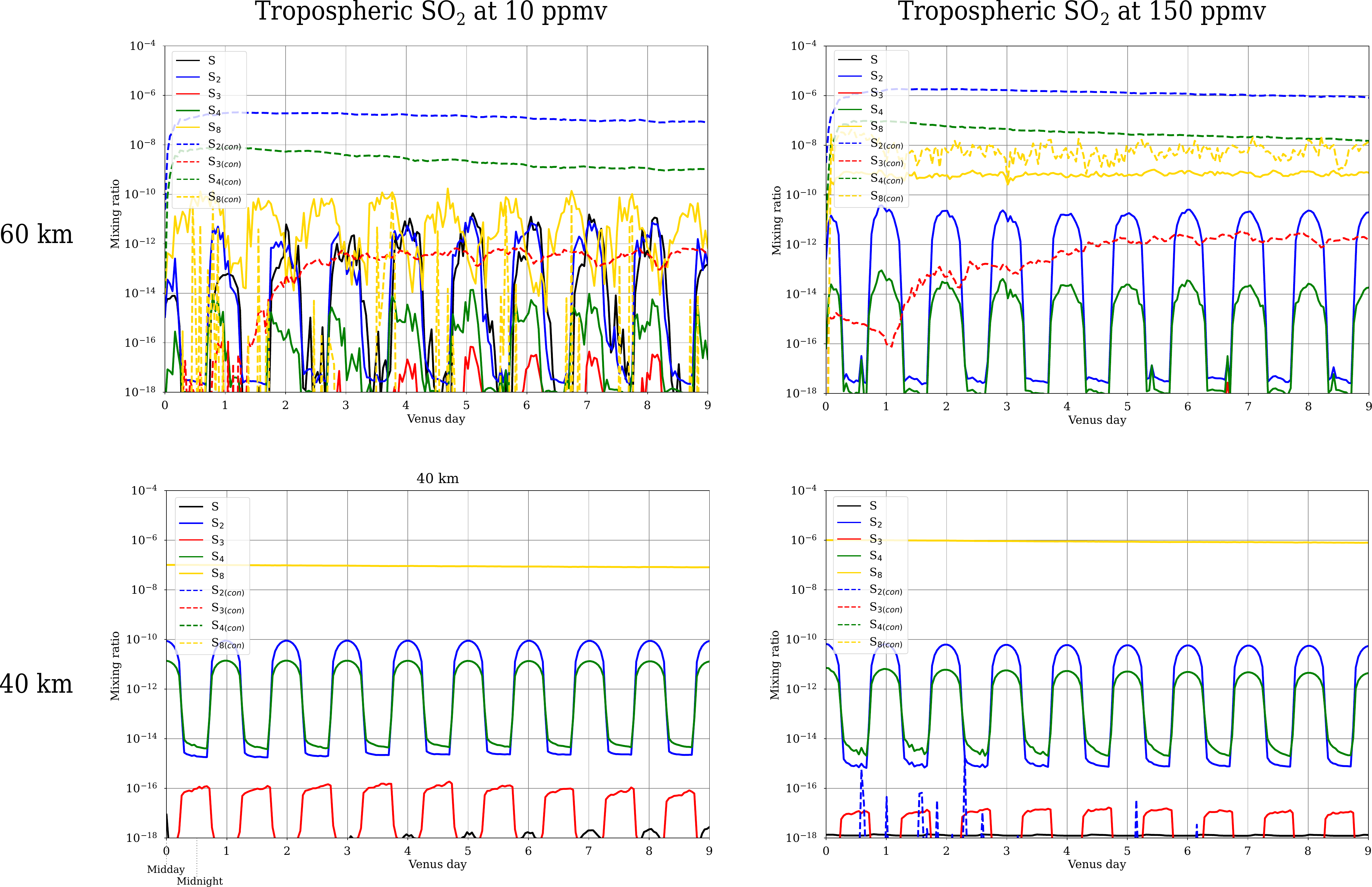}
    \caption{Instantaneous time series of sulfur allotropes mixing ratios in gas phase (solid curves) and condensed phase (dashed curves) at the Equator and longitude 0$^{\circ}$ and at 40 (bottom row) and 60~km (top row) for tropospheric SO$_2$ set to 10~pmmv (left column) and 150~ppmv (right column). On the bottom panel, the condensed phases of sulfur allotropes are below volume mixing ratios of 10$^{-18}$.}
  \label{402}
\end{figure}

The time variability of the different sulfur allotropes mixing ratio at two different altitudes at the Equator and longitude 0$^{\circ}$ is displayed in Figure~\ref{402}. At 60~km, the condensed phase of S$_2$ and S$_4$ are the most abundant, with a quick increase during the first day and then stabilises. Daily variations are noticeable for the gas phases, with strong increases of several orders of magnitude from S to S$_4$ at noon due to photolysis and an increase of S$_8$ at night due to the allotropes' polymerisation. Subsequent to the increase of the gas phase, there is a diurnal variability, but of a lower extent and hardly noticeable in the figure. At 40~km, the condensed phases of sulfur allotropes are below mixing ratios of 10$^{-18}$, and the gas phases are more abundant with S$_8$ around 10$^{-7}$, stable in time. Strong diurnal variabilities are visible for allotropes from 1 to 4. S, S$_2$ and S$_4$ are maximum at noon due to the photolysis, whereas S$_3$ is maximum at night. This is due to the efficient photolysis rate of S$_3$ that will produce S and S$_2$, with a characteristic timescale below one minute (Fig~\ref{232}). S$_3$ is then polymerised at night. Meanwhile, the photolysis of S$_4$ is stronger than for S$_3$, but S$_4$ is the product of S$_8$ photolysis, which is orders of magnitude more abundant. S$_8$ does not show diurnal variability due to its less effective photolysis below the clouds. With a tropospheric SO$_2$ mixing ratio at 150~ppmv, the diurnal variability at 40~km is similar for S$_2$, S$_3$ and S$_4$, with comparable mixing ratio, and S$_8$ has increased by a factor ten, reaching ppmv levels. At 60~km, S$_2$ and S$_4$ condensed phase are also the main polysulfur species with a increase of a factor ten. S$_8$ in condensed and gas phase are now the third and fourth polysulfur species. However, the SO$_2$ mixing ratio at this altitude is higher than observations, and will increase the polysulfur reservoir via photolysis and chemsitry
.

In total, the polysulfur chemistry represents a reservoir of about 0.5~ppmv of sulfur at cloud-top altitudes with a tropospheric SO$_2$ at 10~ppmv, and about 4~ppmv with a tropospheric SO$_2$ at 150~ppmv. At cloud-bottom, there is around 1~ppmv of equivalent sulfur ($\Sigma x[S_x]$ with x=1,2,3,4,8) with the sulfur allotropes, one order of magnitude below \cite{krasnopolskyPhotochemicalModelVenus2012}, and similar to \cite{biersonChemicalCyclingVenusian2020} but with S$_7$ as their dominant species. In the upper clouds, the \cite{pintoSulfurMonoxideDimer2021} models calculated mixing ratios between 10$^{-10}$ and 10$^{-15}$ for the different sulfur allotropes species gas phase. These values are several orders of magnitude above the mixing ratio calculated with the Venus PCM in this study. This difference is explained by the absence of photolysis for sulfur allotropes species with a higher polymerization degree above 2, as well as condensation, in \cite{pintoSulfurMonoxideDimer2021}. 

\begin{figure}[!ht]
  \centering
  \includegraphics[width=10cm]{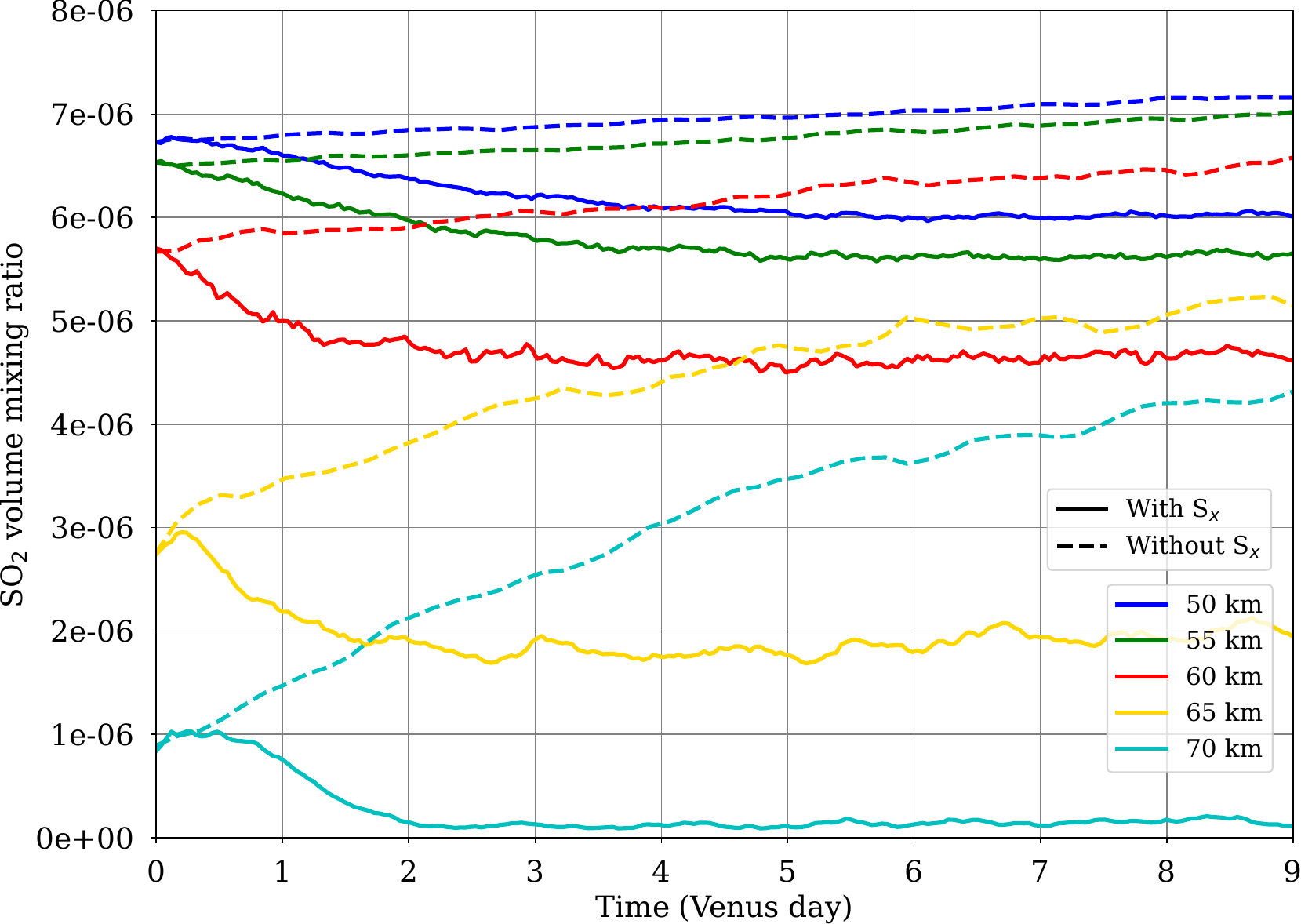}
    \caption{Horizontally-averaged time series of SO$_2$ mixing ratio at five altitudes in the clouds with (solid curves) and without (dashed curves) the sulfur allotrope chemistry described in Section~\ref{Sec:Model1}.}
  \label{403}
\end{figure}

Figure~\ref{403} shows horizontally-averaged time series of SO$_2$ at five altitudes in the clouds. Without polysulfur chemistry, SO$_2$ cloud-top abundance diverges rapidly to unrealistic values of several ppmvs. With the polysulfur chemistry and condensation, the SO$_2$ mixing ratio is stabilised to realistic values in the upper clouds.  With a realistic value of SO$_2$ below the clouds, around 150~ppmv, even with the polysulfur chemistry, SO$_2$ abundance at the cloud-top increases to unrealistic several tens of ppmvs in a few Venus days. In this case, the total polysulfur reservoir is reaching 4~ppmv. Sulfur allotropes are therefore a substantial sulfur sink but not large enough to explain the SO$_2$ vertical gradient in the clouds. S$_x$ chemistry contributes to stabilising the SO$_2$ distribution above the clouds in our simulations.

\section{Sensitivity Studies}
\label{Sec:Sens}
\subsection{Saturation Vapor Pressure}
\label{condensation}

Two main saturation vapour pressure equations are available in the literature \citep{lyonsEstimateEquilibriumSpeciation2008,zahnlePhotolyticHazesAtmosphere2016}, with different hypotheses at temperatures below 410~K. \cite{zahnlePhotolyticHazesAtmosphere2016} considers that the saturation vapour pressure of sulfur allotropes
over the solid is extended to higher temperatures until it
intersects the reported saturation vapour pressure over the liquid, whereas \cite{lyonsEstimateEquilibriumSpeciation2008} is extrapolating the saturation vapour pressure to lower temperatures above solid sulfur allotropes. Simulations were performed with the two sets of equations to quantify the variability related to the saturation vapour pressure uncertainty.

\begin{figure}[!ht]
  \centering
  \includegraphics[width=\textwidth]{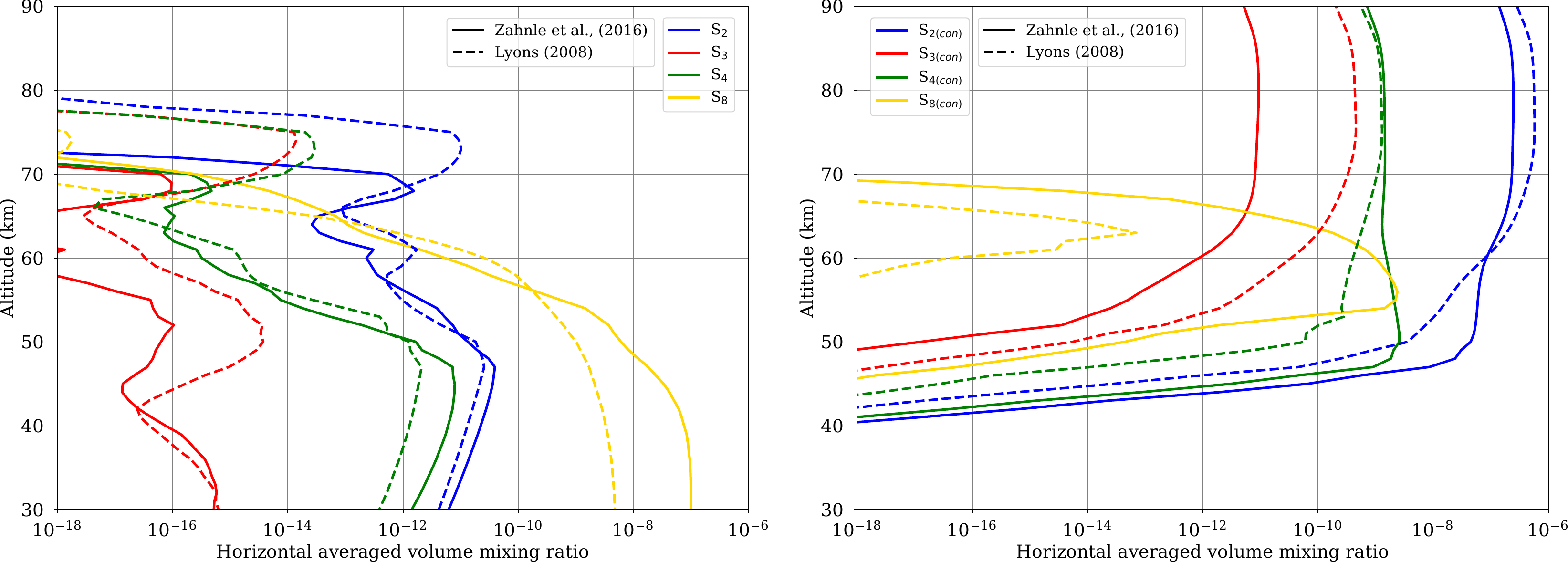}
    \caption{Horizontally-averaged vertical profiles of the sulfur allotropes gas (left) and condensed (right) phase for the \cite{zahnlePhotolyticHazesAtmosphere2016} and \cite{lyonsEstimateEquilibriumSpeciation2008} saturation vapour pressure (respectively plain and dashed curves).}
  \label{411}
\end{figure}

Figure~\ref{411} shows horizontally-averaged vertical profiles of the sulfur allotropes gas (left) and condensed (right) phase for the \cite{zahnlePhotolyticHazesAtmosphere2016} and \cite{lyonsEstimateEquilibriumSpeciation2008} saturation vapour pressures. The \cite{zahnlePhotolyticHazesAtmosphere2016} saturation vapour pressures lead to a lower saturation mixing ratios above 40~km, with a factor of 1000 at 70~km, and increasing above (see Fig~\ref{241}). Therefore, the condensed phase mixing ratios are increased by a factor of ten for S$_2$, a factor of two orders of magnitude for S$_4$ and ten orders of magnitude for S$_8$ at 50~km, and five orders of magnitude between the respective maximum of S$_8$, shifted in altitude. S$_3$ condensed phase is the only species with a lower mixing ratio than with the \cite{lyonsEstimateEquilibriumSpeciation2008} saturation vapour pressures, due to increased condensation of S$_2$ and S$_4$, implicated in its main chemical source reactions. The altitude range of S$_8$ condensed phase location is also affected, with a maximum location about six kilometres lower for \cite{zahnlePhotolyticHazesAtmosphere2016} saturation vapour pressure. The balance between the different polysulfur species is not significantly affected, and among the condensed allotropes species, S$_2$ is the dominant species for both saturation vapour pressure sets of equations.
This uncertainty in sulfur allotropes saturation vapour pressure above 40~km affects marginally the results presented in Section~\ref{Sec:Sx}, except for S$_8$.

\subsection{UV Cross-section}
\label{spectra}

The knowledge of the UV cross-sections for the different sulfur allotropes is not complete. Therefore, recently published updated cross-sections were used in this study \citep{skogUVVisSpectralSimulation2025}. Vertical profile of the photolysis rates of the four considered sulfur allotropes at the Equator at noon are displayed in Figure~\ref{232}.

\begin{figure}[!ht]
  \centering
  \includegraphics[width=\textwidth]{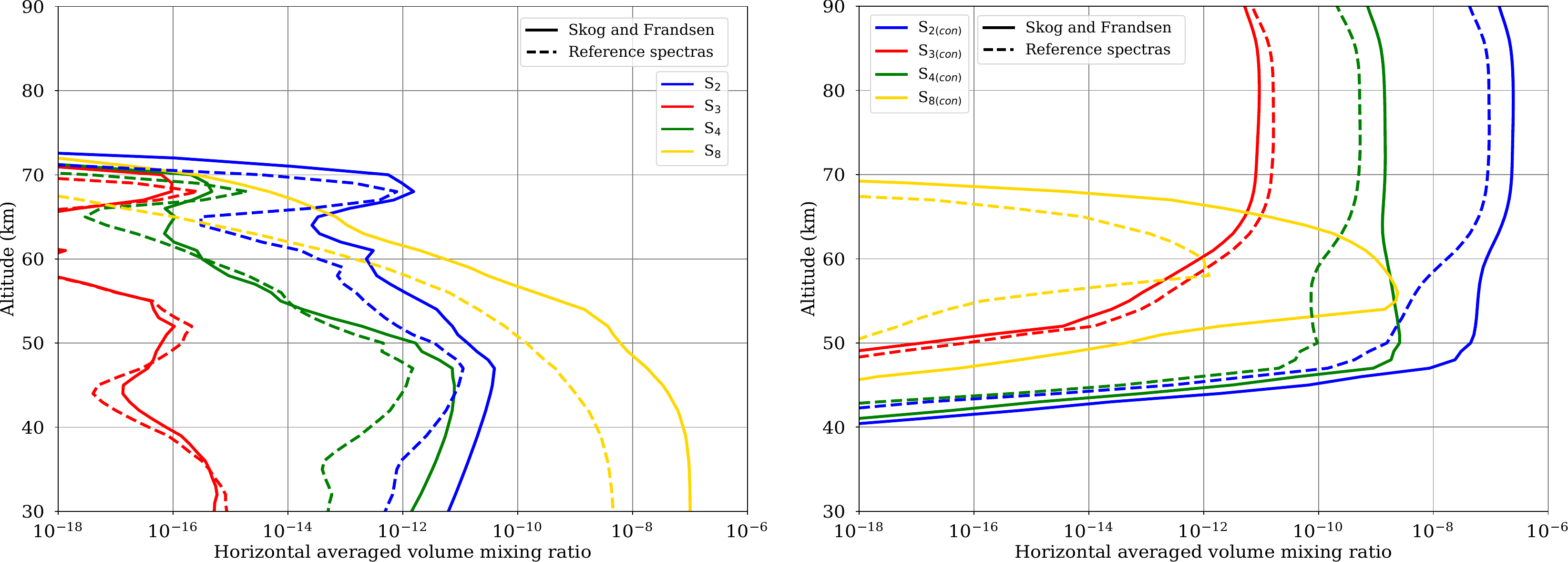}
    \caption{Horizontally-averaged vertical profiles of the sulfur allotropes gas (left) and condensed (right) phase for the reference (dashed curves) and updated (solid curves) cross-section dataset (Figure~\ref{231}).}
  \label{421}
\end{figure}

Figure~\ref{421} shows horizontally-averaged vertical profiles of the sulfur allotropes gas (left) and condensed (right) phase for the two sets of UV cross-sections (Figure~\ref{231}). In the gas phase, the most noticeable difference is for S$_8$. In the new UV cross-section, there is a decrease in amplitude by several orders of magnitude and a shift towards the visible, leading to less efficient photolysis and to an increase in the S$_8$ gas phase of around one order of magnitude. The altitude range where S$_8$ gas is present is also affected, with an increase of a couple of kilometres with the updated cross-sections. Due to this increase, the S$_8$ condensed phase is affected, with an increase of the peaking mixing ratio of almost three orders of magnitude, and an altitude range increased by more than six kilometres. The absence of absorption in the visible range with the updated cross-section for S$_2$ is leading to a general decrease in the photolysis rate and no photolysis below 20~km. Subsequently, the mixing ratio of the condensed phase increases by a factor of ten at 50~km. With the increase in the photolysis rate of S$_8$ and the decrease in the photolysis rate of S$_2$, S$_4$ is strongly affected below the clouds due to the increase in its two main sources. The gas phase mixing ratio is increased by orders of magnitude with the updated cross-section, leading to a similar increase in the condensed phase. S$_3$ gas and condensed phase are hardly affected by the new cross-sections. With the new set of cross-sections, among condensed species, S$_2$ is by far the dominant species, followed by S$_4$ and S$_8$. The update of sulfur allotropes cross sections changes the results for S$_8$ and affects the balance between the different condensed species above in the cloud layer.

\subsection{Polysulfur Particle Radius}
\label{radius}

The particle radius of the sulfur allotropes condensed phase is not known. As a reference, the radius was fixed at 0.1~$\upmu$m for the four considered species. The radius will impact the sedimentation of the species, as well as the radiative properties that are not considered in this study. To quantify the sensitivity of our results on this parameter, several test cases were considered. One simulation was performed with a fixed radius of 1~$\upmu$m for the four considered species. One simulation was performed with a radius that depended on the degree of polymerisation. S$_{2(con)}$ is fixed at 0.1~$\upmu$m, S$_{3(con)}$ at 0.2~$\upmu$m, S$_{4(con)}$ at 0.4~$\upmu$m, and S$_{8(con)}$ at 0.8~$\upmu$m close to experimental values \citep{onischukSurfaceTensionSulfur2016}.

\begin{figure}[!ht]
  \centering
  \includegraphics[width=10cm]{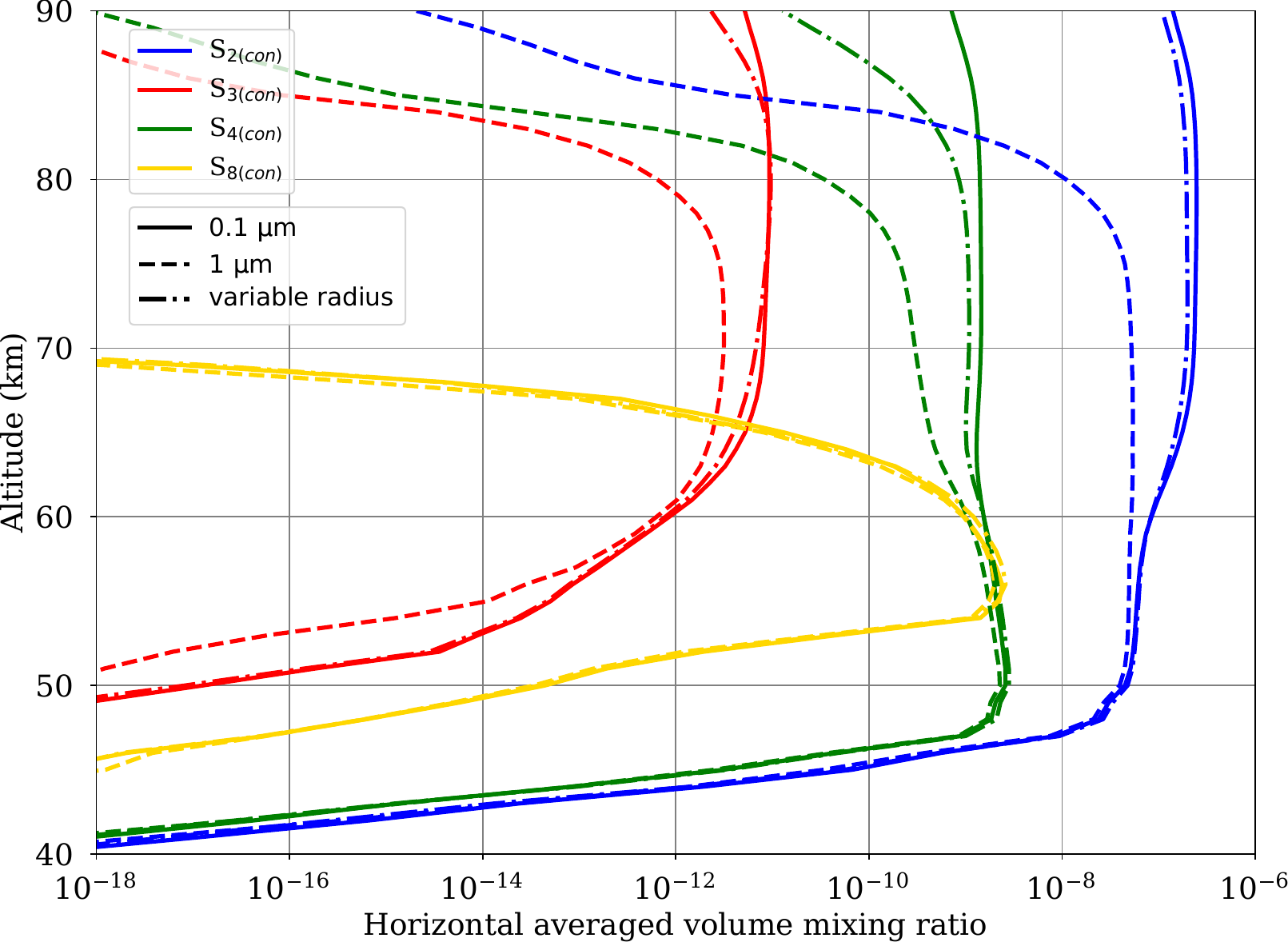}
    \caption{Horizontally-averaged vertical profiles of the sulfur allotropes condensed phase for a common 0.1~$\upmu$m (solid curves) and 1~$\upmu$m (dashed curves) particle radius and a radius depending on the polymerisation degree (dashed dotted curves).}
  \label{441}
\end{figure}

Figure~\ref{441} shows horizontally-averaged vertical profiles of the sulfur allotropes condensed phase for the three different particle radius parametrisations. Below 60~km, the difference in mixing ratio of the condensed phase for the various values is minimal, except for S$_3$ condensed phase decreasing by several orders of magnitude for larger particles. Whereas above, a larger particle radius leads to a significant decrease in the mixing ratio. This is due to the increase in the settling velocity, proportional to the square of the particle radius, and inversely proportional to the viscosity of the atmosphere. At 80~km, particles are sedimenting 34\% faster than at 60~km and 76\% faster than at 50~km for S$_2$ particles (Table~S2). This leads to condensed-rich parcels entering lower altitude regions with higher saturation mixing ratios, engendering a mixing ratio decrease with altitude for the different sulfur allotropes condensed phase for higher particle radius. The balance between the different allotropes remains unchanged. The uncertainty in the sulfur allotropes particle radius marginally affects our conclusions in the cloud layer, but has a significant impact above.

\subsection{S$_8$ Photodissociation Efficiency}
\label{S8}

S$_8$ is in a ring shape configuration. \cite{youngVenusCloudMicrophysics1983} suggested that due to this shape, the photolysis mechanism could be tempered. The first UV photon absorbed would break the ring, and then the linear S$_8$ molecule would either be put back into a ring or split into two S$_4$ molecules by a secondary photon. The quantum yield of S$_8$ in methanol was calculated experimentally \citep{bevilacquaFluorescencePhotochemicalYields1991} at a low value of 0.043 at 266~nm photon absorption, meaning that the ring shape is reducing the efficiency of the photolysis. However, experimental or theoretical data in atmospheric conditions are missing regarding this mechanism, which is essential for achieving a realistic balance between the different polysulfur species. To account for the decreased efficiency of this process, test cases with an S$_8$ photolysis divided by a factor of 10 and 100 were performed, in line with the experimental measurements.

\begin{figure}[!ht]
  \centering
  \includegraphics[width=\textwidth]{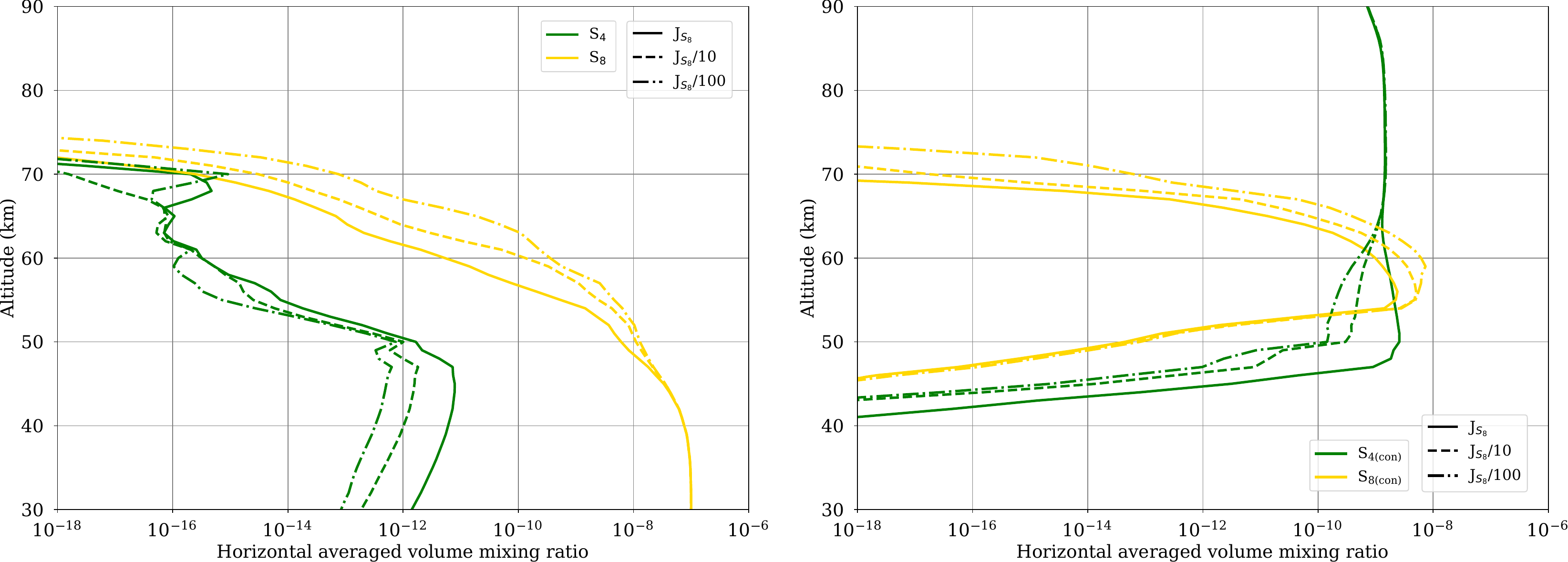}
    \caption{Horizontally-averaged vertical profiles of the sulfur allotropes gas (left) and condensed (right) phase for the calculated photolysis of S$_8$ (solid curves), photolysis of S$_8$ divided by 10 (dashed curves) and photolysis of S$_8$ divided by 100 (dashdotted curves).}
  \label{431}
\end{figure}

Figure~\ref{431} shows the horizontally-averaged vertical profiles of the sulfur allotropes gas (left) and condensed (right) phase for the different S$_8$ photolysis efficiencies. Below 45~km, the difference in S$_8$ gas phase is minimal, whereas S$_4$ decreases as the S$_8$ photolysis efficiency decreases. In the upper cloud, the S$_8$ photolysis divided by a factor of 100 results in an increase between a factor of 2 orders of magnitude and 3 orders of magnitude of S$_8$ gas mixing ratio. This increase is translated into a subsequent increase of the S$_8$ condensed phase by a factor of 3 at the maximum mixing ratio, and an increase of 4~km of the maximum altitude. S$_4$ condensed phase is also affected, with a decrease due to the lower efficiency of the S$_8$ photolysis, around a factor 10 at 50~km. Its altitude range is also affected, with a decrease of several kilometres. More importantly, with this change of the S$_8$ photolysis efficiency, the balance between the different species is modified. At maximum efficiency, S$_4$ and S$_8$ condensed phase are of both around 10$^{-3}$~ppmv in the 50-60~km range. But with a diminished efficiency, S$_8$ condensed phase dominates. The maximum at high latitude (Figure~\ref{311}) is increased by 50\%, and there is the presence of S$_8$ condensed phase at all latitudes between 58 and 61~km with a 10$^{-3}$~ppmv mixing ratio. The uncertainty on the S$_8$ photolysis efficiency affects our conclusions marginally.

\section*{Conclusion}
\label{Sec:Conc}

In this study, we used for the first time a 3D climate model to investigate the climatology of the sulfur allotropes and sulfur hydrides chemistry and their impact on the sulfur cycle. \\
H$_2$S is around ppmv levels from the surface up to the cloud base, with a tropospheric SO$_2$ mixing ratio of both 10 and 150~ppmv. Above, H$_2$S is photolysed and strongly decreases above the clouds. This 1~ppmv value is consistent with SOIR and ground measurements upper limit in the upper clouds. Such a reservoir of sulfur is not negligible in the lower clouds and troposphere, with a value close to ppmv, the third most abundant sulfur-bearing species below the clouds after SO$_2$ and OCS below, and thus the sulfur hydride chemistry needs to be considered to satisfy a realistic chemical network. \\
Polysulfur chemistry represents another substantial sulfur reservoir, but this time in the upper cloud. At cloud top, the sum of equivalent sulfur contained in S$_x$ species reaches an equivalent of 0.5~ppmv with 10~ppmv tropospheric SO$_2$ values, and around four ppmv with realistic 150~ppmv of SO$_2$. 
Compared to the sub-ppmv measured value of SO$_2$ at 70~km, polysulfur chemistry and condensation are a substantial sulfur sink that should be taken into account for future venusian cloud modelling, but not large enough to explain the SO$_2$ vertical gradient in the clouds. \\
Below the clouds, S$_8$ in the gas phase is the most abundant sulfur allotrope. In the cloud layer, the ratio between the different allotropes species changes due to temperature and photolysis. The different species in the gaseous phase dominate up to 50~km, above which the photolysis and condensation are preponderant, decreasing to extremely low values. Sulfur allotropes can therefore not be the unknown absorber at cloud-top altitudes. The gas phase exhibits strong diurnal variability. Most of the allotropes are maximum at noon due to the photolysis of species with a higher degree of polymerisation. But S$_8$ in the clouds and S$_3$ below the clouds are maximum at night, where the polymerisation is more effective. \\
The allotropes in the condensed phase are present, i.e. with mixing ratio above 10$^{-30}$, from 30~km to above 100~km. These species could therefore be implicated in the formation of the haze layer, but S$_8$ condensed phase evaporates strongly below the cloud and cannot be implicated in the lower haze as speculated. This altitude range is consistent for the sulfur allotropes condensed phase to act as cloud condensation nuclei. These condensed phases are the dominant allotropes above 50~km. S$_2$ condensed phase is the most abundant due to its low saturation mixing ratios. S$_8$ condensed phase is only above the ppbv values at high latitudes in the middle and lower clouds. \\

There are several unknown parameters in the modelling of sulfur allotropes and sulfur hydride chemistry. In this study, sensitivity studies were conducted on several key aspects to quantify their impact on the sulfur climatology. Updated UV cross-sections for sulfur allotropes were used. In these cross-sections, the visible range absorption is decreased compared to previous spectras. This crucial change has modified the maximum abundance, altitude range, and equilibrium between the different species. Further theoretical and experimental efforts are needed to model the photolysis of these species realistically.\\
The particle size of condensed sulfur allotropes is not well known. The sensitivity study conducted with radii between 0.1 and 1~$\upmu$m shows that in the cloud layer, the impact is minimal. Above the clouds, the increased particle radii are leading to a significant change. With larger settling velocities due to the lower atmospheric viscosity, the sedimenting parcels are transported in regions with a higher saturation mixing ratio, generating evaporation.\\
S$_8$ photolysis is taken into account for the first time in a Venusian cloud study. It is essential to the equilibrium between the different sulfur allotrope species. Without S$_8$ photolysis, the polymerisation can achieve higher degrees, and S$_8$ condensed phase becomes the secondary condensed sulfur allotrope species. The S$_8$ photolysis efficiency due to its ring shape needs to be quantified both theoretically and experimentally.

Interactions of the sulfur allotropes in the condensed phase with sulfuric acid and water droplets are not considered in this study. Such interaction and its potential role as cloud condensation nuclei should be further investigated.\\
DAVINCI will measure sulfur allotropes and sulfur hydrides from the lower cloud down to the surface, and provide insightful information for the cloud chemistry. However, the role of sulfur allotropes in tropospheric chemistry should be further investigated.

\section*{Acknowledgments}
The authors thank the two anonymous reviewers for their constructive comments. This project has received funding from the European Union's Horizon Europe research and innovation program under the Marie Sk\l odowska-Curie grant agreement 101110489/MuSICA-V. This study benefited from the IPSL Data and Computing Center ESPRI which is supported by CNRS, SU, CNES and Ecole Polytechnique. R.S. acknowledges the support from the Doctoral Programme in Chemistry and Molecular Sciences (CHEMS-DP) at the University of Helsinki. B.N.F. thanks the Carlsberg Foundation (CF22-0754) and the Magnus Ehrnrooth Foundation for fellowship support. Aurélien Stolzenbach acknowledges support from the project PID2021-126365NB-C21, funded by MCIN/AEI/10.13039/501100011033/ and FEDER.

\section*{Open research}
The simulation outputs used to obtain the results in this paper are available in the open online repository \citep{Lefe25data}.

\section*{Conflict of Interest}
The authors declare that they have no conflicts of interest relevant to this study.

\section*{Supporting Information}

\begin{table}
\tiny
\begin{tabular}{lll|lll}
Vertical layer & Pressure (Pa) & Pseudo-altitude (km) & Vertical layer & Pressure (Pa) & Pseudo-altitude (km)\\
\hline
1 & 9.26283500e+06 & 0.0 & 40 & 5.05137085e+02 & 78.09571838378906\\
2 & 9.23998100e+06 & 0.03894679248332977 & 41 & 3.61946991e+02 & 79.55894470214844\\
3 & 9.17871700e+06 & 0.14377687871456146 & 42 & 2.59346191e+02 & 80.99964904785156\\
4 & 9.05096200e+06 & 0.36438044905662537 & 43 & 1.85829849e+02 & 82.40790557861328\\
5 & 8.83362100e+06 & 0.7459183931350708 & 44 & 1.33152618e+02 & 83.77397155761719\\
6 & 8.50837800e+06 & 1.332099437713623 & 45 & 9.54081802e+01 & 85.09246826171875\\
7 & 8.06418600e+06 & 2.164046287536621 & 46 & 6.55746613e+01 & 86.53557586669922\\
8 & 7.49937650e+06 & 3.2794153690338135 & 47 & 3.98794479e+01 & 88.42802429199219\\
9 & 6.82324000e+06 & 4.711222171783447 & 48 & 2.12784443e+01 & 90.80595397949219\\
10 & 6.05635150e+06 & 6.486660003662109 & 49 & 1.02023306e+01 & 93.56942749023438\\
11 & 5.22937800e+06 & 8.62586498260498 & 50 & 4.52917528e+00 & 96.63746643066406\\
12 & 4.37996150e+06 & 11.141559600830078 & 51 & 2.37315726e+00 & 99.09341430664062\\
13 & 3.54840200e+06 & 14.038281440734863 & 52 & 1.44410026e+00 & 100.97420501708984\\
14 & 2.77271325e+06 & 17.310096740722656 & 53 & 8.75891745e-01 & 102.8695068359375\\
15 & 2.08375938e+06 & 20.9365177154541 & 54 & 5.31254172e-01 & 104.78823852539062\\
16 & 1.51368238e+06 & 24.787574768066406 & 55 & 3.22221845e-01 & 106.73812866210938\\
17 & 1.08362362e+06 & 28.586212158203125 & 56 & 1.95437655e-01 & 108.71430206298828\\
18 & 7.75562812e+05 & 32.162296295166016 & 57 & 1.18538864e-01 & 110.7119140625\\
19 & 5.54961812e+05 & 35.52327346801758 & 58 & 7.18974993e-02 & 112.71698760986328\\
20 & 3.97159000e+05 & 38.678489685058594 & 59 & 4.36080061e-02 & 114.65731811523438\\
21 & 2.84416312e+05 & 41.64599609375 & 60 & 2.64496114e-02 & 116.48983764648438\\
22 & 2.03786703e+05 & 44.45757293701172 & 61 & 1.60424821e-02 & 118.23828125\\
23 & 1.46019266e+05 & 47.13932800292969 & 62 & 9.73027665e-03 & 119.92784118652344\\
24 & 1.04627586e+05 & 49.68362045288086 & 63 & 5.90170128e-03 & 121.57347869873047\\
25 & 7.49689531e+04 & 52.069862365722656 & 64 & 3.57956463e-03 & 123.18083953857422\\
26 & 5.37175430e+04 & 54.295509338378906 & 65 & 2.17111432e-03 & 124.76073455810547\\
27 & 3.84903125e+04 & 56.368133544921875 & 66 & 1.31684716e-03 & 126.3230209350586\\
28 & 2.75795215e+04 & 58.3095817565918 & 67 & 7.98709632e-04 & 127.8674087524414\\
29 & 1.97616230e+04 & 60.160831451416016 & 68 & 4.84441407e-04 & 129.39010620117188\\
30 & 1.41598281e+04 & 61.952030181884766 & 69 & 2.93828576e-04 & 130.89646911621094\\
31 & 1.01459521e+04 & 63.69272994995117 & 70 & 1.78216040e-04 & 132.4063720703125\\
32 & 7.26989795e+03 & 65.39546203613281 & 71 & 1.08093605e-04 & 133.9485626220703\\
33 & 5.20911719e+03 & 67.08006286621094 & 72 & 6.55619733e-05 & 135.5504913330078\\
34 & 3.73249634e+03 & 68.76139068603516 & 73 & 3.97653785e-05 & 137.22894287109375\\
35 & 2.67444971e+03 & 70.41968536376953 & 74 & 2.41189246e-05 & 138.98611450195312\\
36 & 1.91632678e+03 & 72.0338363647461 & 75 & 1.46288648e-05 & 140.8113555908203\\
37 & 1.37310779e+03 & 73.6030502319336 & 76 & 8.87284205e-06 & 142.6865997314453\\
38 & 9.83874939e+02 & 75.12689208984375 & 77 & 5.38165432e-06 & 144.5926971435547\\
39 & 7.04976501e+02 & 76.61894989013672 & 78 & 3.26414033e-06 & 146.51504516601562\\
 
\hline
\end{tabular}
\caption{Venus PCM vertical grid}
\label{ST1}
\end{table}

\begin{figure}[!ht]
  \centering
  \includegraphics[width=10cm]{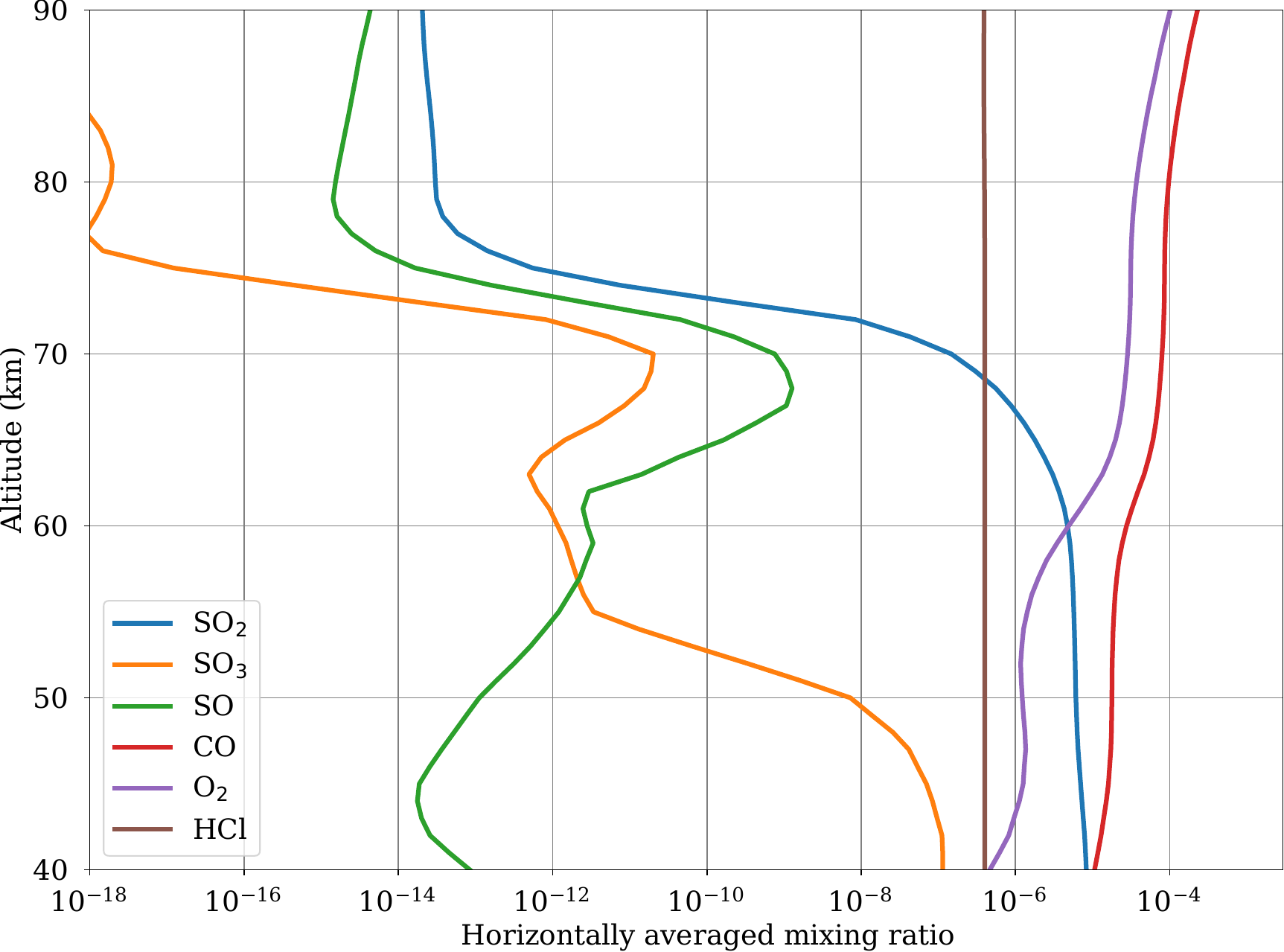}
    \caption{Horizontally-averaged mixing ratio vertical profiles of minor species.}
  \label{S1}
\end{figure}

\subsection*{Updated cross-sections for the sulfur allotropes}

The absorption cross-sections for the polysulfur species were taken from a recent publication \citep{skogUVVisSpectralSimulation2025}, where the UV-Vis spectra were simulated using the nuclear ensemble approximation approach. Geometry optimizations and harmonic vibrational frequency calculations for the S$_2$-S$_4$ structures were performed at the CCSD/aug-cc-pV(T+d)Z level of theory \citep{woonGaussianBasisSets1993,dunningGaussianBasisSets2001}, while the $\omega$B97X-D/aug-cc-pV(T+d)Z \citep{chaiLongrangeCorrectedHybrid2008,chaiSystematicOptimizationLongrange2008} level of theory was used for S$_8$. Spectral simulations were carried out with the Newton-X 2.6 (build 01) code \citep{barbattiNewtonXPlatformNew2022}. A nuclear ensemble of 2000 geometries was created based on a Wigner distribution of calculated harmonic vibrational frequencies. Vertical excitation energies and corresponding oscillator strengths were calculated using EOM-CCSD/aug-cc-pV(T+d)Z for the S$_2$-S$_4$ structures, and using $\omega$B97X-D3/aug-cc-pV(T+d)Z for S$_8$ \citep{linLongRangeCorrectedHybrid2013}. The ten lowest energy vertical excitations were included in the calculations. A 0.1 eV full-width at half-maximum (FWHM) normalized Gaussian function was used to convolute the calculated excitations. All spectral simulations were performed at a temperature of 0~K.

\subsection*{Settling velocity parameter}

As defined in the main text, the terminal velocity $V$ based on the Stokes law \citep{fuchsMechanicsAerosols1964,ackermanPrecipitatingCondensationClouds2001}: 

\begin{equation}
    V = \frac{2\upbeta r^2 g(\uprho_p - \uprho)}{9\upeta}
\end{equation}

With $r$ is the particle radius, $g$ the gravitational acceleration, $\uprho_p$ is the particle density, $\uprho$ is the atmosphere density, $\upeta$ is the viscosity of the atmospheric gas calculated following \citep{johnstonViscositiesCarbonMonoxide1942} and the \cite{jonesDeterminationMolecularFieldsI1924} formalism, and $\upbeta$ is the Cunnigham slip factor given by \citep{fuchsMechanicsAerosols1964}:
\begin{equation}
    \upbeta = 1 + K_n(1.256 + 0.4e^{-1.1/Kn})
\end{equation}

With $K_n$ is the Knudsen number, equal to the ratio of the mean free path $\uplambda$ to the size of radius $r$ of the particle: $K_n$ = $\uplambda$/r

The condensed particle mean free path $\uplambda$ is given by
\begin{equation}
    \uplambda = \frac{k_B T}{\sqrt{2}\pi d}\frac{1}{P}
\end{equation}

With $k_B$ the Boltzmann constant, $T$ and $P$ the temperature and pressure of the gas and
$d$ the molecular diameter.

The density of the allotropes particles is set 1400, 1500, 1600 and 2070 kg/m$^3$ for respectively S$_2$, S$_3$, S$_4$ and S$_8$.

Table~\ref{ST2} summarize the value of the different physical parameters for three altitudes in the clouds and particle radius of 0.1 and 1~$\upmu$m.

\begin{table}[!t]
\tiny
\begin{tabular}{lllllllllll}
\hline
Altitude (km) & T (K) & P (Pa) & $\uprho$ (kg/m$^3$) & $\upeta$ (Pa.s) & r ($\upmu$m) & $\uplambda$ (m) & $K_n$ & $\upbeta$ & V$_{S_2}$ (m/s) & V$_{S_8}$ (m/s) \\
\hline
80 & 195 & 474 & 0.013 & 9.9$\cdot$10$^{-06}$ & 0.1 & 6.39$\cdot$10$^{-18}$ & 6.4$\cdot$10$^{-11}$ & 1.0000000000802585 & 2.7$\cdot$10$^{-06}$ & 4.1$\cdot$10$^{-06}$ \\
80 & 195 & 474 & 0.013 & 9.9$\cdot$10$^{-06}$ & 1 & 6.4$\cdot$10$^{-19}$ & 6.4$\cdot$10$^{-13}$ & 1.0000000000008030 & 2.8$\cdot$10$^{-04}$ & 4.1$\cdot$10$^{-04}$ \\
\hline
70 & 240 & 3960 & 0.086 & 1.20$\cdot$10$^{-05}$ & 0.1 & 9.4$\cdot$10$^{-19}$ & 9.4$\cdot$10$^{-12}$ & 1.0000000000118063 & 2.3$\cdot$10$^{-06}$ & 3.4$\cdot$10$^{-06}$ \\
70 & 240 & 3960 & 0.086 & 1.20$\cdot$10$^{-05}$ & 1 & 9.4$\cdot$10$^{-20}$ & 9.4$\cdot$10$^{-14}$ & 1.0000000000001181 & 2.2$\cdot$10$^{-04}$ & 3.4$\cdot$10$^{-04}$ \\
\hline
60 & 266 & 23837 & 0.47 & 1.33$\cdot$10$^{-05}$ & 0.1 & 1.7$\cdot$10$^{-19}$ & 1.7$\cdot$10$^{-12}$ & 1.0000000000021352 & 2.1$\cdot$10$^{-06}$ & 3.1$\cdot$10$^{-06}$ \\
60 & 266 & 23837 & 0.47 & 1.33$\cdot$10$^{-05}$ & 1. & 1.7$\cdot$10$^{-20}$ & 1.7$\cdot$10$^{-14}$ & 1.0000000000000213 & 2.1$\cdot$10$^{-04}$ & 3.0$\cdot$10$^{-04}$\\
\hline
50 & 352 & 106758 & 1.58 & 1.74$\cdot$10$^{-05}$ & 0.1 & 5.1$\cdot$10$^{-20}$ & 5.1$\cdot$10$^{-13}$ & 1.0000000000006406 & 1.6$\cdot$10$^{-06}$ & 2.3$\cdot$10$^{-06}$ \\
50 & 352 & 106758 & 1.58 & 1.74$\cdot$10$^{-05}$ & 1. & 5.1$\cdot$10$^{-21}$ & 5.1$\cdot$10$^{-15}$ & 1.0000000000000064 & 1.6$\cdot$10$^{-04}$ & 2.3$\cdot$10$^{-04}$ \\
\hline
\end{tabular}
\caption{Settling velocity parameter}
\label{ST2}
\end{table}

\newpage
\bibliographystyle{apalike}
\bibliography{biblio}

\end{document}